\documentclass[reprint,aps,amsmath,amssymb,amsfonts]{revtex4-1}
\usepackage[]{graphicx,color}
\usepackage{braket,bm,comment}
\usepackage{dcolumn}
\begin{document}

\title{Effects of the three-dimensional interplanar coupling on the centrosymmetric skyrmion crystal formation in the frustrated stacked-triangular Heisenberg model}
\author{R. Osamura}
\author{K. Aoyama}
\affiliation{Department of Earth and Space Science, Graduate School of Science, Osaka University, Osaka 560-0043, Japan}
\author{K. Mitsumoto}
\affiliation{Graduate School of Arts and Sciences, The University of Tokyo, Tokyo 153-8902, Japan}
\author{H. Kawamura}
\email[Electronic address: ]{h.kawamura.handai@gmail.com}
\affiliation{Molecular Photoscience Research Center, Kobe University, Kobe, 657-8501, Japan}



\date{\today}

\begin{abstract}
 Effects of the three-dimensional (3D) interplanar coupling on centrosymmetric skyrmion crystal (SkX) formation is investigated via extensive Monte Carlo simulations on the frustrated isotropic Heisenberg model on a stacked-triangular lattice in both cases of the ferromagnetic (F) and the antiferromagnetic (AF) nearest-neighbor interplanar coupling $J_{1c}$. The SkX phase is stabilized at finite fields and at finite temperatures for both F and AF $J_{1c}$, although it is destabilized by modestly weak AF $J_{1c}$. The magnetic phase diagram of the 3D short-range model is more or less similar to those of the 2D short-range model and of the 2D long-range RKKY model. We find that an intriguing phenomenon of replica-symmetry breaking, popular in glass physics and recently identified in the SkX phase of the 3D long-range RKKY model [K. Mitsumoto and H. Kawamura, Phys. Rev. B {\bf 104}, 184432 (2021)], does not arise in the 3D short-range model, suggesting that the long-range nature of interaction might be necessary to realize the RSB in centrosymmetric SkX state.
\end{abstract}


\maketitle

\section{Introduction}

 In recent years, there arises a lot of interest in topology-related sciences including condensed-matter physics. This is primarily because objects possessing a nontrivial topology, i.e., topological objects, are often protected from various perturbations or disturbances from their surroundings owing to its nontrivial topological properties, i.e., topological protection. Topology both in wavevector ($k$) and real spaces have attracted much interest. Topological objects in real space often appear as defects, spatial textures or nonlinear excitations. In recent years, ``skyrmion'', a swirling noncoplanar spin texture characterized by an integer topological charge whose constituent spin directions wrap a sphere in spin space, has attracted much attention in the fields of magnetism and spintronics \cite{NagaosaTokura-review,Fert1-review,Fert2-review,KanazawaTokura-review,TokuraKanazawa-review,Kawamura-review}.

 Skyrmion texture in condensed-matter physics was first recognized by Belavin and Polyakov in ferromagnetic Heisenberg model in two dimensions as a special ``metastable state'', a topologically stable excited state with a finite excitation energy above the ferromagnetic ground state \cite{Polyakov}. After some time, it was revealed that skyrmion could be stabilized even in thermal equilibrium state as a periodic array called the skyrmion crystal (SkX) in a certain magnetically ordered state, directly accessible experimentally \cite{Bogdanov1,Bogdanov2,Pfleiderer,Neubauer,Yu1,Yu2}. Indeed, the triple-$q$ nature of the SkX and the topological Hall effect arising from the quantum Berry phase effect were experimentally observed, which were regarded as characteristics of the SkX state. At an earlier stage, the SkX state was discussed for non-centrosymmetric magnets as induced by the antisymmetric Dzyaloshinskii-Moriya (DM) interaction \cite{Bogdanov1,Bogdanov2,Pfleiderer,Neubauer,Yu1,Yu2}.

 In 2012, it was theoretically proposed  by Okubo, Chung and Kawamura that the ``symmetric'' SkX is also possible in certain class of frustrated centrosymmetric magnets without the DM interaction \cite{OkuboChungKawamura}. It was suggested that, while the frustration-induced centrosymmetric SkX was expected to exhibit the triple-$q$ spin structure and the topological Hall effect similarly to the DM-induced non-centrosymmetric SkX, the size of constituent skyrmion could be an order of magnitude smaller than that of the DM skyrmion. In an ideal highly-symmetric situation, both skyrmion and anti-skyrmion of mutually opposite signs of the topological charge, or of the scalar chirality, might equally be possible. Furthermore, a random domain state consisting of short-range SkX and anti-SkX domains called the $Z$ state emerges, leading to unique and rich electromagnetic responses \cite{OkuboChungKawamura}.

 In Ref. \cite{OkuboChungKawamura}, the SkX was identified 
in a simplified model, i.e., the frustrated $J_1$-$J_3$ (or $J_1$-$J_2$) isotropic Heisenberg model on a two-dimensional (2D) triangular lattice as a triple-$q$ state stabilized by magnetic fields and thermal fluctuations. Subsequent experiment successfully observed the SkX for centrosymmetric triangular-lattice metallic magnet, e.g., Gd$_2$PdSi$_3$, accompanied by the pronounced topological Hall effect \cite{Kurumaji,Hirschberger2020PRL}. Recent Monte Carlo (MC) simulation indicated that the SkX could also be stabilized in the standard RKKY system with only the bilinear interaction modelling weak-coupling metals, where the oscillating nature of the RKKY interaction bears frustration \cite{MitsumotoKawamura2021, MitsumotoKawamura2022}.

  Of course, real material possesses various perturbative interactions not taken into account in a simplified model of Ref. \cite{OkuboChungKawamura}, e.g., the three-dimensionality (interplanar coupling), the magnetic anisotropy, and quantum fluctuations, etc. Among them, the effect of magnetic anisotropy has been studied rather extensively, and turned out to play an important role in the SkX formation \cite{LeonovMostovoy,HayamiLinBatista2016,Kawamura2024}. By contrast, relatively little studies have been made on the effects of the three-dimensionality (interplanar coupling). Since the bulk SkX-hosting magnets are in fact three-dimensional (3D) systems experimentally, which inevitably possess finite amount of interplanar coupling, it is also important to clarify the effect of the interplanar coupling on the SkX formation. 

 Lin and Batista numerically studied the centrosymmetric SkX formation in the frustrated classical Heisenberg on a 3D hexagonal (stacked-triangular) lattice with moderate easy-axis single-ion anisotropy in both cases of unfrustrated and frustrated interplanar couplings \cite{LinBatista3D}. 
 It was observed that the SkX states took various forms along the interplanar direction depending on the type of the interplanar coupling. While the triangular crystal of skyrmion tubes running along the magnetic-field direction was stabilized for the ferromagnetic nearest-neighbor interplanar coupling, more complicated 3D skyrmion structures were also observed for other cases, including the SkX consisting of skyrmion tubes tilted away from the magnetic-field axis, and even the fcc- and the hcp-type 3D skyrmion arrangements \cite{LinBatista3D}. 

 A MC study by Mitsumoto and Kawamura on the centrosymmetric SkX formation of the fully isotropic RKKY Heisenberg model on a stacked-triangular lattice revealed that the SkX state of the 3D long-range RKKY model accompanied a peculiar ordering phenomenon of replica-symmetry breaking (RSB) \cite{MitsumotoKawamura2021}, which is familiar in glass systems  \cite{Mezard,FischerHertz-review,KawamuraTaniguchi} but rather rare in regular systems. In the RSB SkX state, the triple-$q$ SkX was macroscopically degenerate with the single-$q$ spiral state. The situation is in sharp contrast to the SkX state in the 2D $J_1-J_3$ ($J_1-J_2$) model \cite{OkuboChungKawamura} or even to the 2D RKKY model \cite{MitsumotoKawamura2022} where no such RSB was observed.

 In the presence of the RSB, the ordered state in real space consists of macroscopic domains, not only of SkX of positive and negative chiralities (chiral domains), but also of single-$q$ spirals running along three equivalent crystallographic directions of the triangular lattice. Recall that the triple-$q$ SkX state and the single-$q$ spiral state are not related by any symmetry operation of the Hamiltonian. 
It was argued that the observed RSB phenomena were made possible due to the heavy degeneracy between the competing ordered states, and might suggest a possible close analogy of SkX physics to glass physics \cite{Mezard,FischerHertz-review,KawamuraTaniguchi}. 

 Under such circumstances, we wish to investigate in the present paper the effects of the interplanar 3D coupling on the centrosymmetric SkX formation via the systematic MC study on the ordering properties and the magnetic phase diagram of the frustrated $J_1-J_3-J_{1c}$ isotropic Heisenberg model on a 3D stacked-triangular lattice for both cases of ferromagnetic and antiferromagnetic nearest-neighbor interplanar coupling $J_{1c}$.

 Our motivation is twofold: For one, we wish to clarify the stability of the SkX state in the isotropic system. In view of the current experimental situation that the centrosymmetric SkX experimentally identified so far are almost all Gd$^{3+}$ (or Eu$^{2+}$) magnets possessing rather weak magnetic anisotropy \cite{Kawamura-review,Kurumaji,Hirschberger2020PRL,Hirschberger2019,KhanhSeki2020,Takagi2022}, understanding the isotropic limit of the SkX-formation problem would be important. Even in the simplest case of ferromagnetic nearest-neighbor $J_{1c}$, whether the SkX state remains stable or not seems not so trivial, if one recalls the fact that the SkX of the fully isotropic Heisenberg model in 2D with $J_{1c}=0$ was stabilized by thermal fluctuations \cite{OkuboChungKawamura}, and that the fluctuation effect generally tends to be reduced in 3D than in 2D. 

 In Ref. \cite{LinBatista3D}, in the presence of both the easy-axis magnetic anisotropy and the interplanar coupling, the SkX state was found to be stabilized down to zero temperature. Indeed, recent theoretical studies have clarified that the SkX state is stabilized down to zero temperature in the presence of the easy-axis magnetic anisotropy even in 2D \cite{LeonovMostovoy,HayamiLinBatista2016,Kawamura2024}. Yet, the fate of the SkX state in the fully isotropic Heisenberg model in 3D is not necessarily clear. Hence, we wish to clarify first the fate of the SkX state in 3D in the presence of the interplanar coupling. 

 For the other, we wish to clarify the RSB phenomena recently observed in the 3D long-range RKKY Heisenberg model \cite{MitsumotoKawamura2021} exists or not in the 3D Heisenberg model with the short-range couplings. Though understanding the conditions of the RSB has long remained challenging in glass physics, general wisdom obtained via extensive studies is that the RSB is more likely  to occur in higher spatial dimensions and for longer-range interactions \cite{Mezard,FischerHertz-review,KawamuraTaniguchi}. The observation that the RSB occurs in the 3D RKKY model, but not in the 2D RKKY model \cite{MitsumotoKawamura2022} nor in the 2D short-range model \cite{OkuboChungKawamura}, is certainly consistent with such general tendency. It remains to be seen in 3D whether the RSB, already established in the 3D long-range RKKY model, exists or not in the short-range model. Such knowledge would cast further light on the conditions of the occurrence of the RSB in centrosymmetric SkX systems. 

 Via extensive MC simulations, we find that for the frustrated short-range $J_1-J_3-J_{1c}$ Heisenberg model on a 3D stacked triangular lattice the SkX state is stabilized at finite fields ($H$) and at finite temperatures ($T$), for both cases of ferromagnetic and antiferromagnetic $J_{1c}$. The $T$-$H$ phase diagram is more or less similar to those of the 2D models, though the stability range of the SkX state is considerably reduced for the antiferromagnetic $J_{1c}$. We find that the RSB, which arises in the 3D long-range RKKY model, does not occur in the 3D short-range $J_1-J_3-J_{1c}$ model. The result suggests that both the three-dimensionality and the long-range interaction is necessary to realize the RSB in centrosymmetric SkX states.

 The present paper is organized as follows. In \S II, we explain our model and the computation method employed. In \S III, we study the ordering properties and the magnetic phase diagram of the model with the ferromagnetic nearest-neighbor interplanar coupling. In \S IV,  the ordering properties and the magnetic phase diagram of the model with the antiferromagnetic nearest-neighbor interplanar coupling are studied; moderately weak interplanar coupling in \S IV-1, and even weaker interplanar coupling in \S IV-2. Finally, \S V is devoted to summary and discussion.

\section{The model and the method}

 We consider the frustrated isotropic classical Heisenberg model on a 3D stacked-triangular (or a simple-hexagonal) lattice, where each triangular-lattice layer forms a direct on-top stack on an adjacent triangular layer. The interactions in the triangular layer are taken to be the competing ferromagnetic $J_1>0$ and antiferromagnetic $J_3<0$, while the interplanar interaction $J_{1c}$ is assumed to work only between nearest neighbors along the $z$-axis, i.e., the $J_1-J_3-J_{1c}$ model. The Hamiltonian is given by
\begin{eqnarray}
{\cal H} = &-& J_1 \sum_{<ij>} {\bm S}_i\cdot {\bm S}_j  -J_3 \sum_{<<ij>>} {\bm S}_i\cdot {\bm S}_j \nonumber \\
  &-& J_{1c} \sum_{<ij>_c} {\bm S}_i\cdot {\bm S}_j   - H\sum_i S_{iz} ,
\end{eqnarray}
where ${\bm S}_i=(S_{ix},S_{iy},S_{iz})$ is the classical Heisenberg spin of unit length with $|{\bm S}_i|=1$ located at $i$-th site on a 3D stacked-triangular lattice, and the magnetic field is applied along the $S_z$ direction. The summations in the first and the second terms represent the sum over the intraplanar nearest-neighbor and third-neighbor pairs on triangular layers, respectively, while the one in the third term represents the sum over the nearest-neighbor pairs along the interplanar stacking direction. The intraplanar interaction ratio is set to $J_1/J_3=-1/3$, taken to be the same as the one employed in the 2D $J_1-J_3$ model studied in Ref. \cite{OkuboChungKawamura}. The interplanar interaction could be either ferromagnetic ($J_{1c}>0$) or antiferromgnetic ($J_{1c}<0$). 

 We study the ordering properties of the model by means of extensive MC simulations. The total number of spins is $N=L\times L\times L_z$, where $L$ is the linear size of the triangular layer, while $L_z=rL$ is the linear size along the stacking ($z$) direction. In examining the size dependence of the MC data, we vary $L$ for a given fixed aspect ratio $r$. Periodic boundary conditions are imposed in all three directions.  

 MC simulation based on the standard heat-bath method combined with the over-relaxation method is employed. Both the $T$-sweep at constant-$H$ and the $H$-sweep at constant-$T$ runs are made. In addition, at relatively high-$T$ region, fully equilibrated $T$-exchange runs are also made. Single MC step consists one heat-bath updating followed by 10 over-relaxation sweeps. At a given $T$ and $H$, total $2\times 10^5$ MC steps per spin (MCS) are generated, and the first half is discarded for thermalization.

\section{Ferromagnetic interplanar coupling}

 We  begin with the case of the $J_1-J_3-J_{1c}$ 3D Heisenberg model with the ferromagnetic nearest-neighbor interplanar coupling. In case of the ferromagnetic $J_{1c}>0$, the ground-state spin structure in zero field is a single-$q$ spiral in a wide parameter range of $0\leq J_1/|J_3| < 4$ \cite{OkuboChungKawamura}, characterized by the ordering wavevector ${\bm q}^*=(q_x^*, q_y^*, q_z^*)$  running along the nearest-neighbor direction on the triangular layer with $q_z^*=0$. Reflecting the $C_3$ lattice-rotation symmetry, there are three equivalent ordering wavevectors ${\bm q}_1^*,{\bm q}_2^*$ and ${\bm q}_3^*$, each related by the $C_3$ lattice rotation. The absolute value of ${\bm q}^*$ is given by $|\bm{q}^\ast|=\frac{2}{d}\cos^{-1}\left[\frac{1}{4}\left(1+\sqrt{1-\frac{2J_1}{J_3}}\right)\right]$, $d$ being the lattice constant which is taken to be unity in the following \cite{OkuboChungKawamura}. Indeed, this parameter range of $J_1/|J_3|$ covers continuously from the 120$^\circ$ structure at $J_1/|J_3|\rightarrow 0$ to the long-wavelength limit ${\bm q}^*\rightarrow 0$ at $J_1/|J_3|\rightarrow 4$ where the skyrmion size becomes infinite corresponding to the continuum limit.  For the strength of the interplanar coupling $J_{1c}$, we mainly study the case of $J_{1c}/|J_3|=1/15$.

 The ordering of the model is studied by MC simulations  as a function of the temperature $T$ and the magnetic field $H$. The lattice size is taken to be $L\times L\times \frac{2}{3}L$, i.e., $L_z=\frac{2}{3}L$ ($r=\frac{2}{3}$). Note that, in studying the size dependence of physical quantities, $L$ is varied with the aspect ratio ($r$) being fixed to a common value, $r=\frac{2}{3}$ here.

\begin{figure}[t]
	\centering
	\begin{tabular}{c}
		\begin{minipage}{\hsize}
			\includegraphics[width=\hsize]{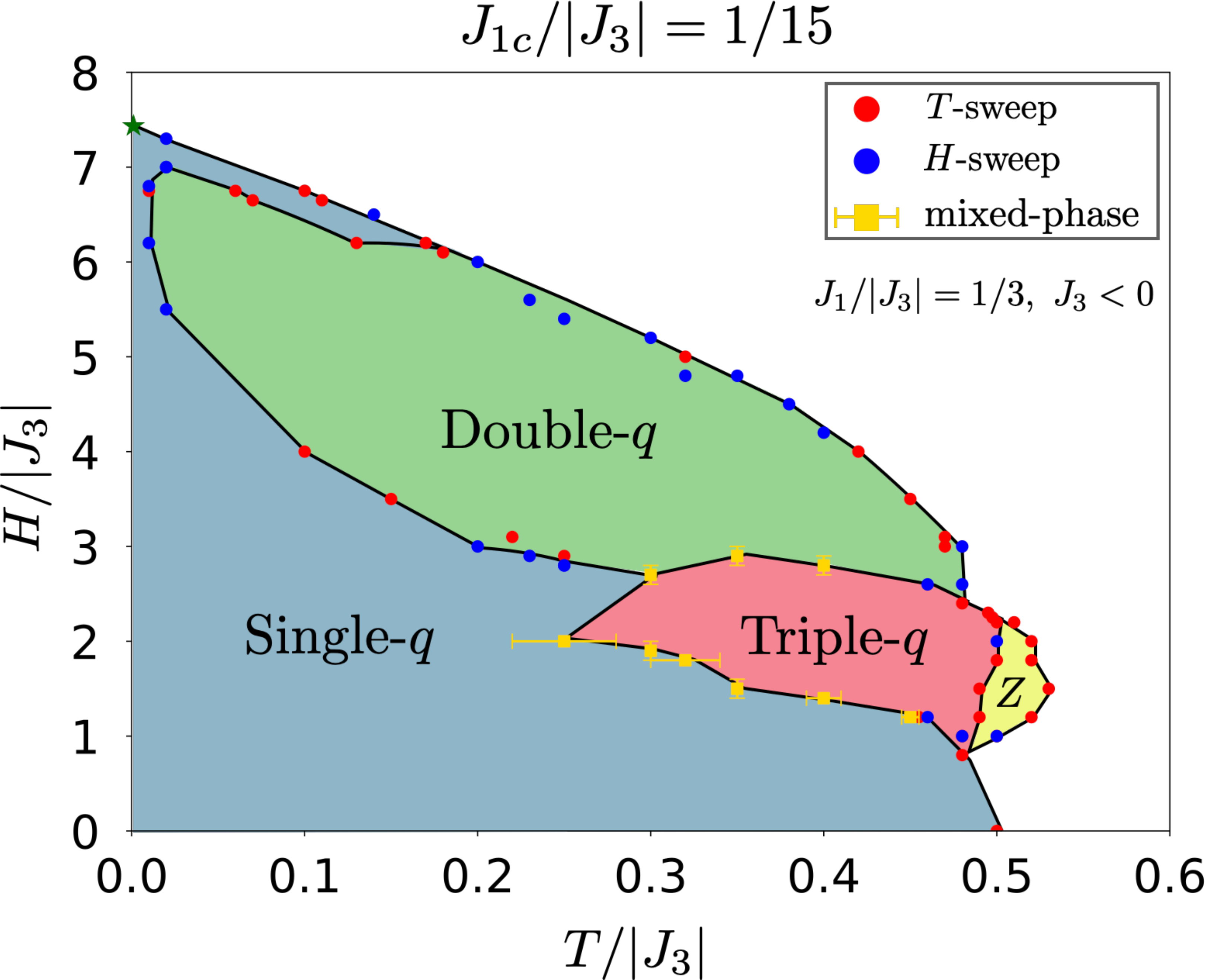}	
		\end{minipage}
	\end{tabular}
	\caption{
The temperature ($T$) versus magnetic field ($H$) phase diagram of the frustrated $J_1-J_3-J_{1c}$ classical Heisenberg model on a 3D stacked-triangular lattice with the ferromagnetic nearest-neighbor interplanar coupling of $J_{1c}/|J_3|=1/15$ as determined by MC simulations, where $J_1$ and $J_3$ are the ferromagnetic nearest-neighbor and the antiferromagnetic third-neighbor intraplanar couplings with $J_1/|J_3|=1/3$. Phase boundaries are determined both by $T$- and $H$-sweeps, the mixed-phase method also being employed. In the mixed-phase method, the stability of the competing phases is numerically determined by performing MC simulations starting from special initial states consisting of the coexistence of the two phases in question, and by monitoring during the subsequent MC evolutions which phase expands over the other \cite{mixed phase}.  
}
	\label{3DFPD}
\end{figure}

 We first show in Fig. 1 the $T$-$H$ phase diagram of the model determined by MC simulations. The obtained 3D phase diagram turns out to be rather similar to the 2D phase diagram reported in Ref. \cite{OkuboChungKawamura}. In particular, the SkX phase persists at intermediate fields and at finite temperatures, together with the $Z$ phase located right to the SkX phase, and is transformed into the single-$q$ transverse conical-spiral phase on further lowering $T$ toward $T=0$. 

  In Fig. 2, we show the $T$-dependence of various physical quantities computed at a particular field of $H/|J_3|=1.2$. At this field, on decreasing $T$ from higher $T$, the system visits four distinct phases exhibiting three transitions as para $\rightarrow$ $Z$ $\rightarrow$ triple-$q$ SkX $\rightarrow$ single-$q$ conical-spiral phases. In our $T$-sweep runs, both $T$-annealing runs and $T$-exchange runs are employed, the latter being limited to relatively high-$T$ range, only down to the middle of the SkX phase. Thus, the data shown in Figs. 2 are taken by the $T$-exchange runs in the higher-$T$ range, which are connected to the ones taken by the $T$-annealing runs in the lower-$T$ range. The entire data shown in Figs. 2, however, seem to be well thermalized. As shown in Fig. 2(a), the specific heat exhibits clear anomalies at the three transition points, although the finite-size effect is still considerable.
\begin{figure}[t]
	\centering
	\begin{tabular}{c}
		\begin{minipage}{\hsize}
			\includegraphics[width=\hsize]{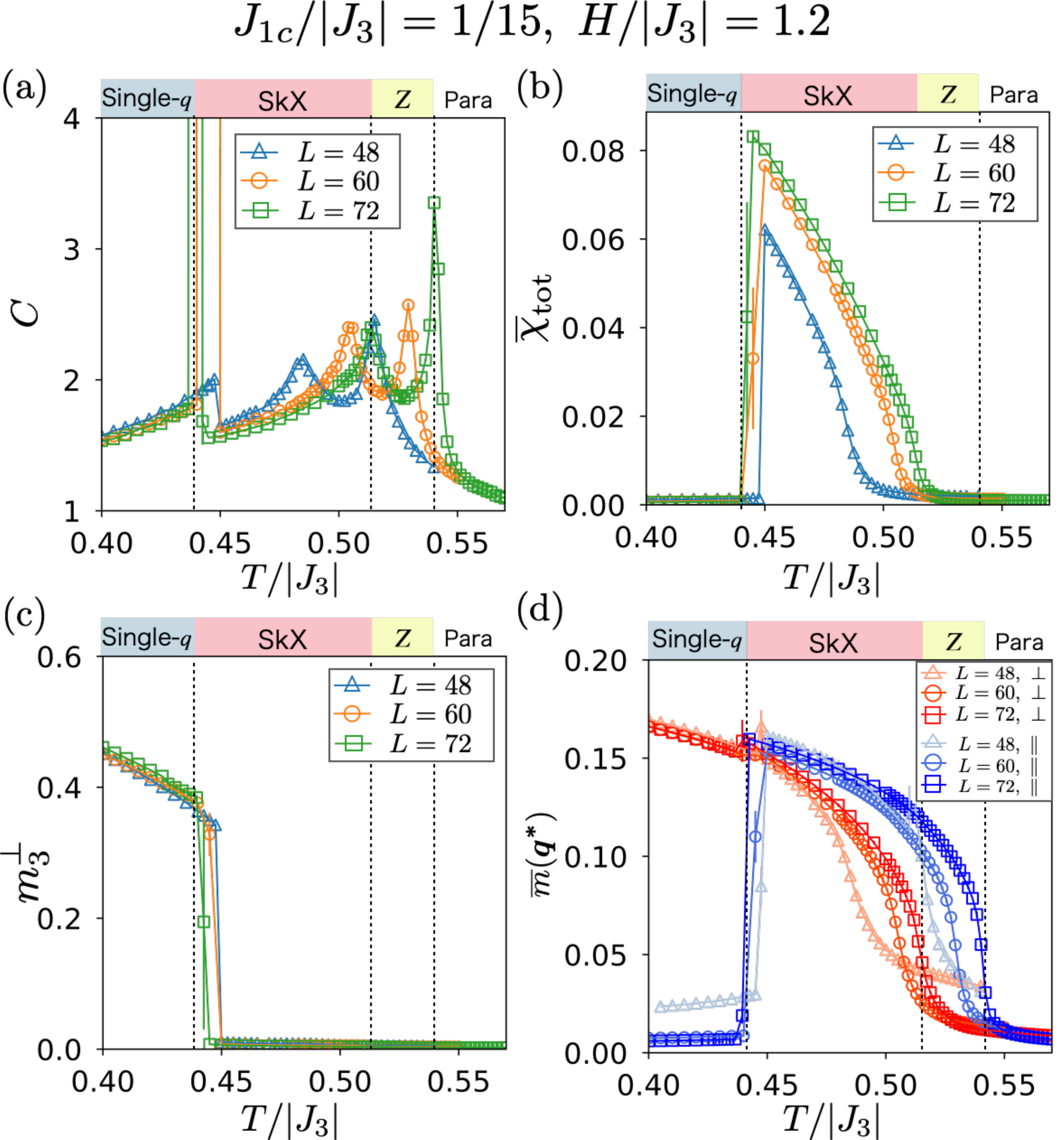}	
		\end{minipage}
	\end{tabular}
	\caption{
The temperature and size dependence of (a) the specific heat, (b) the total scalar chirality, (c) the transverse component of the lattice $C_3$-symmetry-breaking parameter, and (d) the magnetic order parameters associated with the ordering wavevectors ${\bm q}^*$ both for the transverse ($\perp$) and longitudinal ($\parallel$) spin components, at a field $H/|J_3|=1.2$ of the frustrated $J_1-J_3-J_{1c}$ Heisenberg model on a stacked-triangular lattice with the ferromagnetic interplanar coupling of $J_{1c}/|J_3|=1/15$. The lattice sizes are $L\times L\times \frac{2}{3}L$ with $L=48, 60$ and 72. 
	}
\end{figure}

 An important quantity characterizing the SkX state might be the scalar spin chirality, which is directly related to the topological Hall effect via the quantum Berry phase \cite{Ye, Ohgushi, TataraKawamura}. Local scalar chirality might be defined for the three neighboring Heisenberg spins at the sites $i,j$ and $k$ as $\chi_{ijk}={\bm S}_i\cdot ({\bm S}_j\times {\bm S}_k$), and the total scalar chirality is defined by

\begin{eqnarray}
\chi_{\rm tot} &=& \frac{1}{2N}\left(\sum_{\bigtriangleup} \chi_{\bigtriangleup} + \sum_{\bigtriangledown} \chi_{\bigtriangledown}\right) , \label{chitot} \\
\overline\chi_{{\rm tot}} &=& \sqrt{\langle \chi_{{\rm tot}}^2 \rangle} ,
\end{eqnarray}
where $\chi_{\bigtriangleup} (\chi_{\bigtriangledown})$ represents the local scalar chirality $\chi_{ijk}$ for the three spins on an upward (downward) elementary triangle in the triangular-lattice layer, $\langle \cdots \rangle$ means the thermal average, and the summation is taken over all upward and downward elementary triangles on a stacked-triangular lattice covering the entire triangular layers. 

 The $T$-dependence of the total scalar chirality is shown in Fig. 2(b). As can be seen from the figure, $\overline \chi_{{\rm tot}}$ takes a nonzero value only in the SkX phase leading to the net topological Hall effect, while it vanishes in the $Z$ and the single-$q$ phases.

 Fig. 2(c) exhibits the $T$-dependence of the $C_3$ lattice-rotation symmetry-breaking parameter associated with the spin-transverse ($S_x,S_y$) components, $m^\perp_3$, defined by
\begin{eqnarray}
m_3^\perp=\langle |\bm{m}_3^\perp|\rangle ,\ \ \ \ \bm{m}_3^\perp=\sum_{\mu=1}^3 \epsilon_\mu^\perp \hat e_\mu, \\
\epsilon_\mu^\perp = \frac{1}{N} \sum_{i,\mu} (S_i^x S_{i+\mu}^x + S_i^y S_{i+\mu}^y), 
\end{eqnarray}
where $\hat e_1=(0,1)$, $\hat e_2=(-\frac{\sqrt{3}}{2},-\frac{1}{2})$ and $\hat e_3=(\frac{\sqrt{3}}{2},-\frac{1}{2})$, the sum over $i$ is taken over all sites on the 3D stacked-triangular lattice, and the sum over $\mu$ denotes three nearest-neighbor directions of the triangular lattice. Note that in the present model the ordering wavevector ${\bm q}^*$ runs along the nearest-neighbor directions on the triangular layer. Likewise, one can introduce the $C_3$ symmetry-breaking parameter associated with the spin-longitudinal component, $m_3^\parallel$, by
\begin{eqnarray}
m_3^\parallel=\langle |\bm{m}_3^\parallel| \rangle ,\ \ \ \ \bm{m}_3^\parallel=\sum_{\mu=1}^3 \epsilon_\mu^\parallel \hat e_\mu, \\
\epsilon_\mu^\parallel = \frac{1}{N} \sum_{i,\mu} S_i^z S_{i+\mu}^z, 
\end{eqnarray}

 As can be seen from Fig. 2(c), $m^\perp_3$ becomes nonzero only in the single-$q$ phase, indicating a spontaneous $Z_3$-symmetry breaking occurring there. The result is consistent with the observation that the $Z$ and SkX phases keep the $C_3$ lattice-rotation symmetry, while the single-$q$ phase spontaneously breaks it. 

 Let us define the perpendicular and parallel spin structure factors at the 3D wavevector ${\bm q}=(q_x,q_y,q_z)$ by, %
\begin{eqnarray}
S^\perp (\bm{q}) &=& \frac{1}{N} \left\langle \sum_{\alpha = x,y} \left| \sum_{i=1}^N S_{i\alpha} e^{-i \bm{q}\cdot \bm{r}_i}\right|^2  \right\rangle, \label{insta_perp}  \\
S^\parallel (\bm{q}) &=& \frac{1}{N} \left\langle \left| \sum_{i=1}^N S_{iz} e^{-i \bm{q}\cdot \bm{r}_i} \right|^2 \right\rangle, \label{insta_para}
\end{eqnarray}
respectively.

For the ferromagnetic $J_{1c}$, since $J_{1c}$ favors the $q_z=0$ order which is also favored by applied magnetic fields, the 3D magnetic order is expected at $q_z=0$. If we decompose the 3D wavevector ${\bm q}$ into the transverse and the longitudinal components as ${\bm q}=({\bm q}_{xy},q_z)$ with ${\bm q}_{xy}=(q_x,q_y)$, the ordering wavevectors are $\pm {\bm q}_1^*$,  $\pm {\bm q}_2^*$ and  $\pm {\bm q}_3^*$ where  ${\bm q}_j^*=({\bm q}_{j,xy}^* ,0)$ ($j=1,2,3$).

 As the magnetic order parameter, we introduce the mean order-parameter amplitude $\overline m({\bm q}^*)$ via the associated spin structure factor $S({\bm q})$, for each case of the spin-transverse ($S_x,S_y$) components (perpendicular to the field) and the spin-longitudinal $S_z$ component (parallel with the field) by 
\begin{eqnarray}
\overline m^\perp({\bm q}^*) = \sqrt{\frac{1}{3N} \left( S^\perp ({\bm q}_1^*)+S^\perp ({\bm q}_2^*)+S^\perp({\bm q}_3^*) \right) } , \\
\overline m^\parallel({\bm q}^*) = \sqrt{\frac{1}{3N} \left( S^\parallel({\bm q}_1^*)+S^\parallel({\bm q}_2^*)+S^\parallel({\bm q}_3^*) \right) } .
\end{eqnarray}

 The computed $T$-dependence of the order-parameter amplitudes $\overline m^\perp({\bm q}^*)$ and $\overline m^\parallel({\bm q}^*)$ is shown in Fig. 2(d). As can be seen from the figure, the SkX state has both $\overline m^\perp({\bm q}^*)>0$ and $\overline m^\parallel({\bm q}^*)>0$, the $Z$ state has $\overline m^\perp({\bm q}^*)=0$ and $\overline m^\parallel({\bm q}^*)>0$, while the single-$q$ spiral state has $\overline m^\perp({\bm q}^*)>0$ and $\overline m^\parallel({\bm q}^*)=0$.


%
%
\begin{figure}[t]
	\centering
	\begin{tabular}{c}
		\begin{minipage}{\hsize}
			\includegraphics[width=0.89\hsize, keepaspectratio]{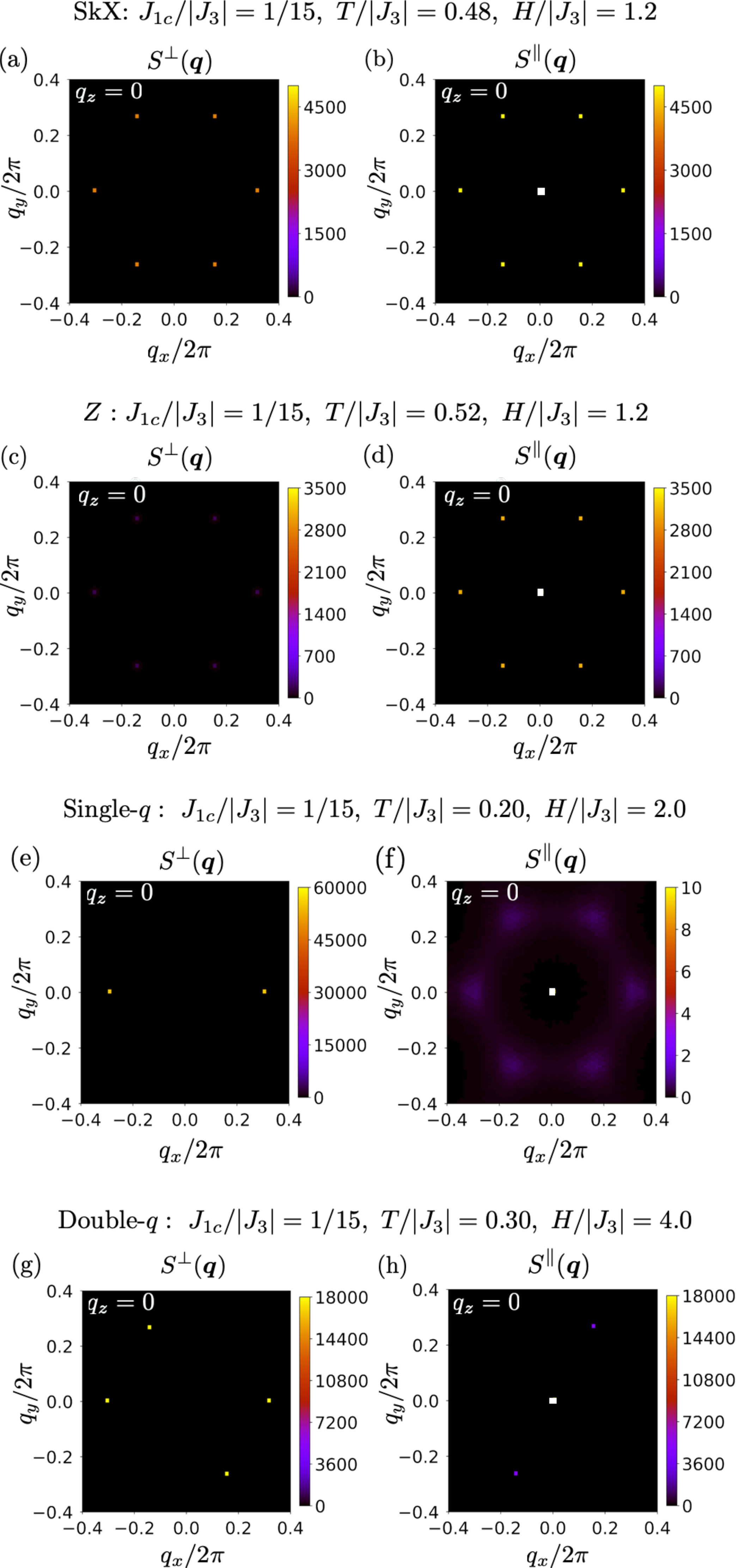}	
		\end{minipage}
	\end{tabular}
	\caption{
Perpendicular and parallel spin structure factors $S^\perp (\bm{q})$ and $S^\parallel (\bm{q})$ in the $(q_x,q_y)$ plane with $q_z=0$, for (a,b) the triple-$q$ SkX phase at $T/|J_3|=0.48$ and $H/|J_3|=1.2$, for (c,d) the $Z$ phase at $T/|J_3|=0.52$ and $H/|J_3|=1.2$, for (e,f) the single-$q$ conical-spiral phase at $T/|J_3|=0.20$ and $H/|J_3|=2.0$, and for (g,h) the double-$q$ phase at $T/|J_3|=0.30$ and $H/|J_3|=4.0$, of the $J_1-J_3-J_{1c}$ model with the ferromagnetic interplanar coupling of $J_{1c}/|J_3|=1/15$. The lattice size is $72\times 72\times 48$ ($L=72$). The data are taken by the $T$-annealing runs.
	}
	\label{3DFPD}
\end{figure}

 In Figs. 3, typical perpendicular and parallel spin structure factors $S^\perp (\bm{q})$ and $S^\parallel (\bm{q})$ are shown in the $(q_x,q_y)$ plane with $q_z=0$, 
 for various ordered phases realized in the magnetic phase diagram, i.e., (a,b) the triple-$q$ SkX phase, (c,d) the $Z$ phase, (e,f) the single-$q$ conical-spiral phase, and (g,h) the double-$q$ phase. As can be seen from Figs. 3 (a, b), the triple-$q$ SkX state, which consists of the superposition of three vertical spirals, gives the $C_3$-symmetric triple-$q$ patterns of Bragg peaks both in $S^\perp (\bm{q})$ and $S^\parallel (\bm{q})$.  In contrast to the 2D $S({\bm q})$ where the peaks consist of quasi-Bragg peaks with power-law spin correlations, the $S({\bm q})$ peaks here should be true Bragg peaks in 3D.

 In the $Z$ state, as can be seen from Figs. 3 (c, d), while the $S^\parallel (\bm{q})$ peaks are sharp Bragg peaks of the triple-$q$ character, $S^\perp (\bm{q})$ exhibits broader peaks corresponding to the short-range order only.

 In the single-$q$ conical-spiral state, As can be seen from Figs. 3 (e, f), $S^\perp (\bm{q})$ exhibits a pair of sharp Bragg peaks, whereas $S^\parallel (\bm{q})$ exhibits only broader peaks corresponding to the short-range order. Such features of $S({\bm q})$ are consistent with the transverse conical-spiral ordered state.

 In the double-$q$ state, as can be seen from Figs. 3 (g, h), $S^\perp (\bm{q})$ exhibits two pairs of sharp Bragg peaks spontaneously breaking the lattice $C_3$ symmetry, while $S^\parallel (\bm{q})$ exhibits one pair of sharp Bragg peaks at the complementary positions. In the notation of Refs. \cite{Kawamura-review,Kawamura2024}, the state might be described as ($2q, 1q$) state, where $m$ and $n$ in ($mq,nq$) represent the number of the strongest Bragg peaks in $S^\perp (\bm{q})$ and $S^\parallel (\bm{q})$, respectively.  The observed features of $S^\perp (\bm{q})$ and $S^\parallel (\bm{q})$ are common with those of the 2D model \cite{OkuboChungKawamura}, although the sharp Bragg-like peaks should be true Bragg peaks in 3D, in contrast to the quasi-Bragg peaks in 2D.
 
 Typical real-space spin configurations of (a) the SkX state, of (b) the $Z$ state, of (c) the single-$q$ state, and of (d) the double-$q$ state are shown in Fig. 4 for three successive triangular layers, from layer 1 to layer 3. The color represents the spin $S_z$ component, while the arrow represents the direction of the spin-transverse ($S_x,S_y$) components. To reduce the thermal noise, short-time averaging of 50 MCS are made. One can see from the figure that, in the SkX state, the skyrmion core forms the triangular superlattice on the atomic (spin) triangular lattice, where the vortex-like swirling spin patterns around the skyrmion core are clearly visible, which are vertically stacked on top of each other forming the skyrmion tube as illustrated in Fig. 5(a).

 In the $Z$ state, which is adiabatically isomorphic to the collinear triple-$q$ state as can be seen from Fig. 3(b), the spin-transverse ($S_x,S_y$) components get much reduced due to their short-range-order character,  while the spin-longitudinal $S_z$ component still exhibits a clear triangular superlattice structure, as can be seen from Fig. 4(b). 

 In the single-$q$ state, the transverse conical spiral runs along one of the nearest-neighbor directions ${\bm q}_1^*,{\bm q}_2^*$ and ${\bm q}_3^*$ by breaking the lattice $C_3$ symmetry, as shown in Fig. 4(c).

 In the double-$q$ state, the doulbe-$q$ spiral is formed in the transverse components consisting of the two ordering wavevectors by breaking the lattice $C_3$ symmetry, while the sinusoidal (spin-density-wave-type) order running along the complementary direction is formed in the longitudinal component, as shown in Fig. 4(d). We note that those real-space spin configurations are fully consistent with the ${\bm q}$-space spin structure factors shown in Fig.3.

\begin{figure}[t]
	\centering
	\begin{tabular}{c}
		\begin{minipage}{\hsize}
			\includegraphics[width=\hsize]{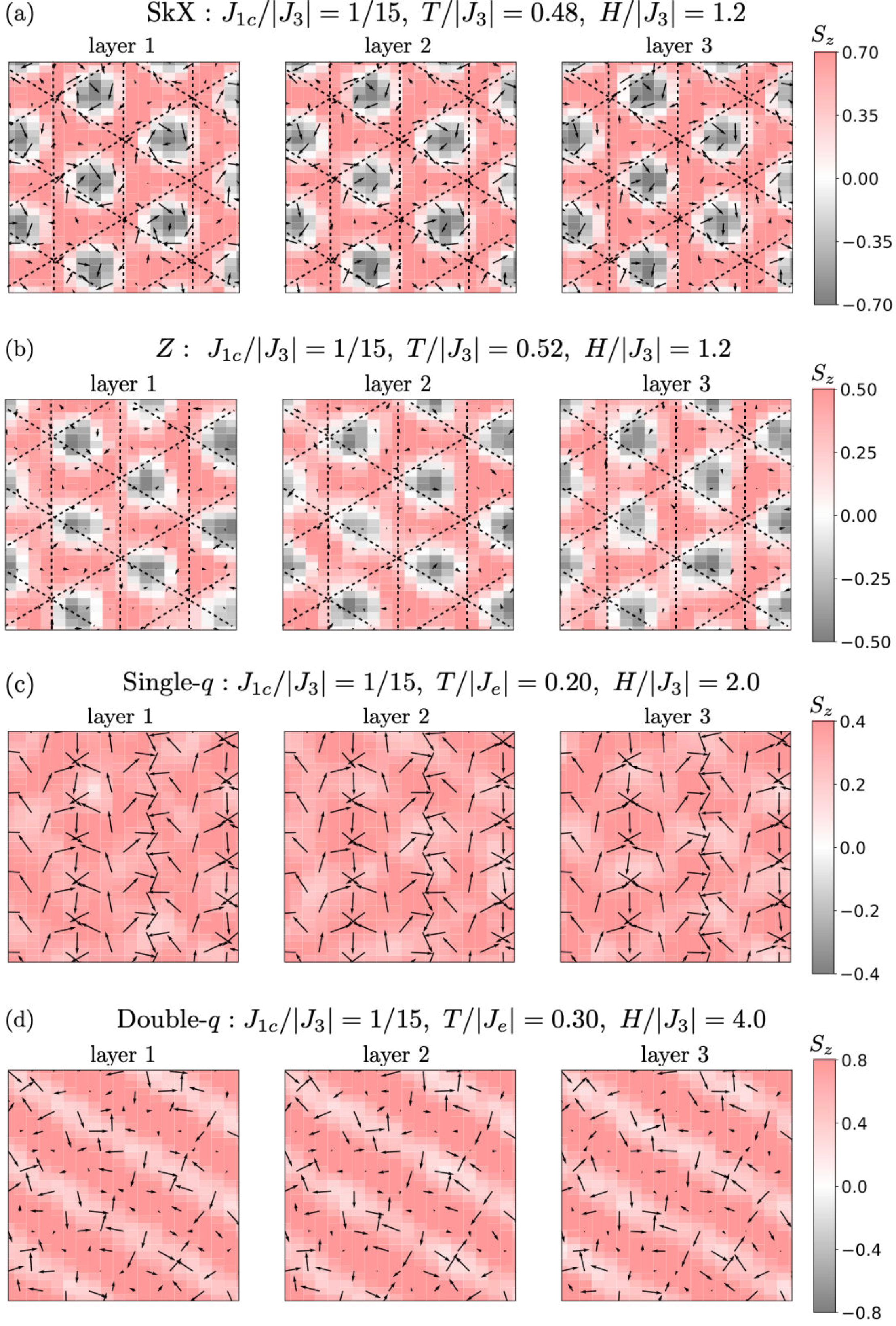}	
		\end{minipage}
	\end{tabular}
	\caption{
Typical real-space spin configurations in (a) the SkX phase at  $T/|J_3|=0.48$ and $H/|J_3|=1.2$, in (b) the $Z$ phase at $T/|J_3|=0.52$ and $H/|J_3|=1.2$, in (c) the single-$q$ phase at $T/|J_3|=0.2$ and $H/|J_3|=2.0$, and in (d) the double-$q$ phase at $T/|J_3|=0.30$ and $H/|J_3|=4.0$, for three successive triangular layers, from layer 1 to layer 3, of the $J_1-J_3-J_{1c}$ model with the ferromagnetic interplanar coupling of $J_{1c}/|J_3|=1/15$. The color represents the spin $S_z$ component, while the arrow represents the direction of the spin-transverse ($S_x,S_y$) components. To reduce the thermal noise, short-time averaging of 50 MCS is made, while the figure represents a part of the $72\times 72\times 48$ ($L=72$) lattice with a common $xy$ section among layers 1-3.  The data are taken by the $T$-annealing runs. 
	}
	\label{3DFPD}
\end{figure}

 The color plots of typical real-space local scalar chirality configurations of (a) the SkX and of (b) the anti-SkX states in the SkX phase are shown in Fig. 6 for three successive triangular layers, together with those of (c) the $Z$ state. One can see that, in the SkX phase, the scalar chiralities are uniformly ordered either to negative (SkX state) or positive (anti-SkX state) value. In contrast, in the $Z$ phase, each triangular layer forms a random domain state consisting of finite-size SkX and anti-SkX domains, with a vanishing net total scalar chirality even for each triangular layer. The stacking pattern of such random-domain states in each layer look also random without long-range correlations along the stacking ($z$) direction. 
\begin{figure}[t]
	\centering
	\begin{tabular}{c}
		\begin{minipage}{\hsize}
			\includegraphics[width=\hsize]{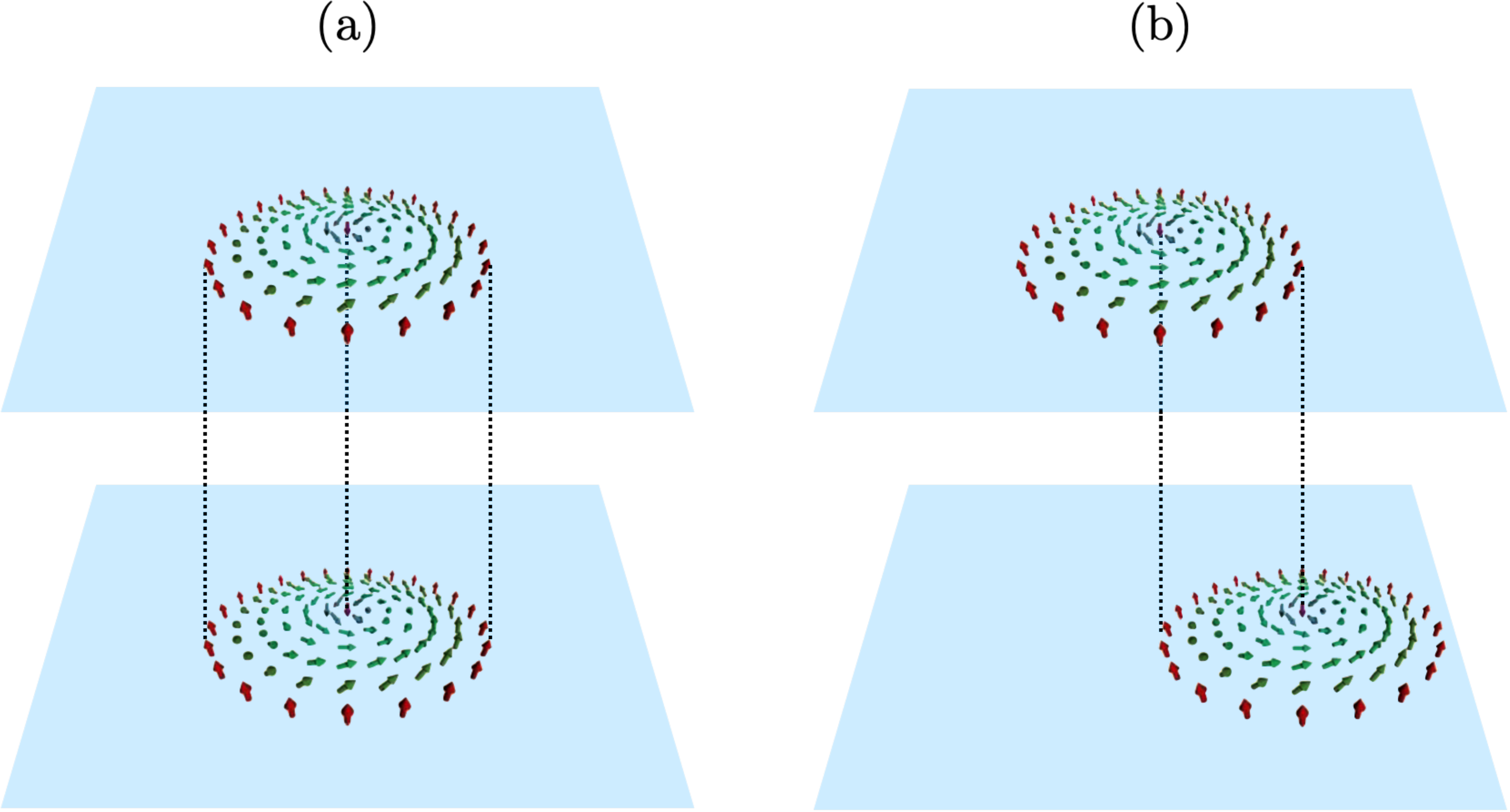}	
		\end{minipage}
	\end{tabular}
	\caption{
Schematic illustration of the manner of the interplanar stacking of the SkX's for two adjacent triangular layers. (a)  Direct on-top stacking realized in the ferromagnetic interplanar coupling $J_{1c}>0$, and (b) slided stacking realized in the antiferromagnetic interplanar coupling $J_{1c}<0$.
	}
	\label{3DFPD}
\end{figure}

 Next, we move to the question of the possible RSB in the SkX phase. As mentioned in \S I, an intriguing RSB phenomenon was observed in the 3D long-range RKKY Heisenberg model on a stacked-triangular lattice \cite{MitsumotoKawamura2021}, while it was not observed in the 2D long-range RKKY Heisenberg model on the triangular lattice \cite{MitsumotoKawamura2022}, nor in the 2D short-range $J_1-J_3$ Heisenberg model on the triangular lattice \cite{OkuboChungKawamura}.  The observed RSB SkX state of the 3D RKKY model consists of macroscopic coexistence of the triple-$q$ SkX state and the single-$q$ conical-spiral state. 

 In Refs. \cite{MitsumotoKawamura2022} and \cite{MitsumotoKawamura2021}, the presence/absence of RSB in the SkX state was probed by investigating the distribution of the total scalar chirality $P(\chi_{{\rm tot}})$, and of the staggered scalar chirality $P(\chi_{\rm stg})$. This is because the total scalar chirality becomes nonzero in the SkX state but vanishes in the single-$q$ conical-spiral state, whereas the staggered scalar chirality becomes nonzero in the single-$q$ conical-spiral state but vanishes in the SkX state. The staggered scalar chirality $\chi_{\rm stg}$ is defined by
\begin{align}
\chi_{\rm stg} &= \frac{1}{2N}\left(\sum_{\bigtriangleup} \chi_{\bigtriangleup} - \sum_{\bigtriangledown} \chi_{\bigtriangledown}\right) .
\end{align}

 The distribution of the total scalar chirality $P(\chi_{{\rm tot}})$, and that of the staggered scalar chirality $P(\chi_{\rm stg})$, computed by the fully equilibrated $T$-exchange runs, are shown in Figs. 7(a) and 7(b), respectively, in the SkX state at $T/|J_3|=0.50$ and $H/|J_3|=1.2$. As can be seen from Fig. 7(a), $P(\chi_{{\rm tot}})$ exhibits two symmetric peaks at nonzero $\chi_{{\rm tot}}=\pm \chi_{{\rm tot}}^*$, which grow and sharpen with increasing the system size $L$, but does not exhibit any appreciable central peak at $\chi_{{\rm tot}}=0$ corresponding to the single-$q$ conical-spiral state.

 This is in sharp contrast to $P(\chi_{{\rm tot}})$ of the 3D RKKY model exhibiting the RSB, which shows three peaks growing with $L$, two symmetric side peaks corresponding to the SkX and anti-SkX states, and one central peak corresponding to the single-$q$ conical-spiral state, signaling the macroscopic coexistence of the triple-$q$ SkX state and the single-$q$ conical-spiral state \cite{MitsumotoKawamura2021}. 
\begin{figure}[t]	\centering
	\begin{tabular}{c}
		\begin{minipage}{\hsize}
			\includegraphics[width=\hsize]{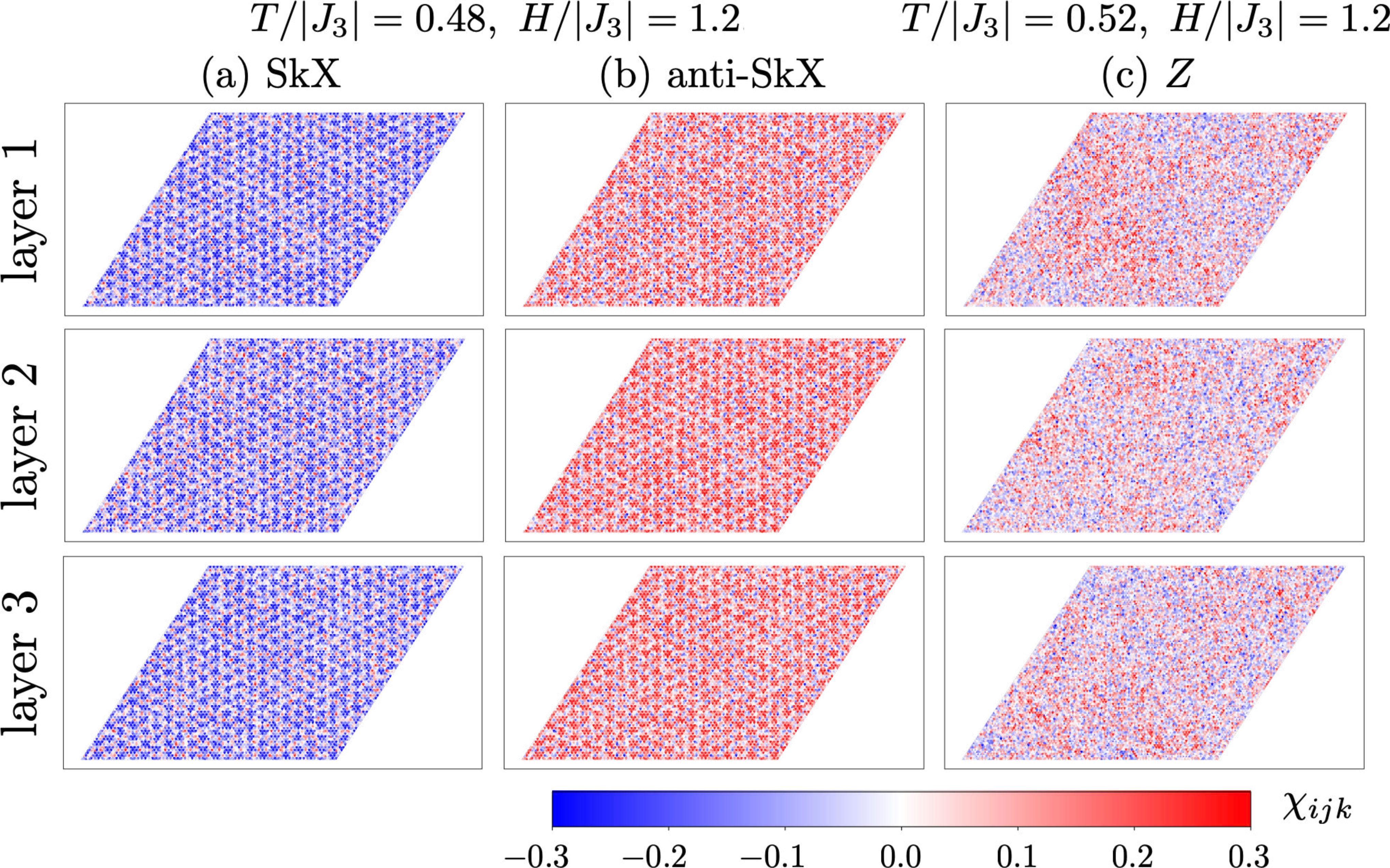}
		\end{minipage}
	\end{tabular}
	\caption{
 Color plots of typical real-space local scalar chirality configurations of (a) the SkX state and (b) the anti-SkX state in the SkX phase  at $T/|J_3|=0.48$ and $H/|J_3|=1.2$, and of (c) the $Z$ state at $T/|J_3|=0.52$ and $H/|J_3|=1.2$, for three successive triangular layers, from layer 1 to layer 3, of the $J_1-J_3-J_{1c}$ model with the ferromagnetic interplanar coupling of $J_{1c}/|J_3|=1/15$. To reduce the thermal noise, short-time  averaging of 50 MCS is made. The figure represents a $72\times 72$ triangular sheet of the 3D stacked-triangular lattice of the size $72\times 72\times 48$. The data are taken by the $T$-annealing runs.
	}
	\label{3DFPD}
\end{figure}

 As can be seen from Fig. 7(b), $P(\chi_{{\rm stg}})$ exhibits only a single central peak at $\chi_{{\rm stg}}=0$ corresponding to the SkX or the anti-SkX state, which grows and sharpens with increasing $L$, but no appreciable peak at other nonzero $\chi_{{\rm stg}}=\pm \chi_{{\rm stg}}^*$ corresponding to the single-$q$ conical-spiral states with nonzero staggered scalar chiralities of mutually opposite signs, in sharp contrast to the three-peak structure characteristic of the RSB observed in $P(\chi_{{\rm stg}})$ of the 3D RKKY model \cite{MitsumotoKawamura2021}.  Thus, the computed $P(\chi_{{\rm stg}})$ again consists of the contribution only from the triple-$q$ SkX (and its $Z_2$-symmetry partner, the anti-SkX) state.

 Based on these observations, we conclude that the SkX state of the present 3D short-range model does not exhibit the RSB, in sharp contrast to the 3D RKKY model. Our observation then suggests the importance of the long-range nature of the RKKY interaction for the occurrence of the RSB, together with the three-dimensionality.

 In order to check the possible dependence of the phase structure on the strength of the ferromagnetic nearest-neighbor interplanar coupling $J_{1c}$, we also studied the case of stronger $J_{1c}/|J_3|=1/3$. It turns out that all qualitative features of the $T$-$H$ phase diagram and the properties of each ordered phase are essentially the same as those of $J_{1c}/|J_3|=1/15$ shown above. Hence, the basic features of the magnetic ordering including the SkX formation seem to be rather robust against the variation of the ferromagnetic interplanar coupling strength as long as it is purely ferromagnetic.
\begin{figure}[t]
	\centering
	\begin{tabular}{c}
		\begin{minipage}{\hsize}
			\includegraphics[width=\hsize]{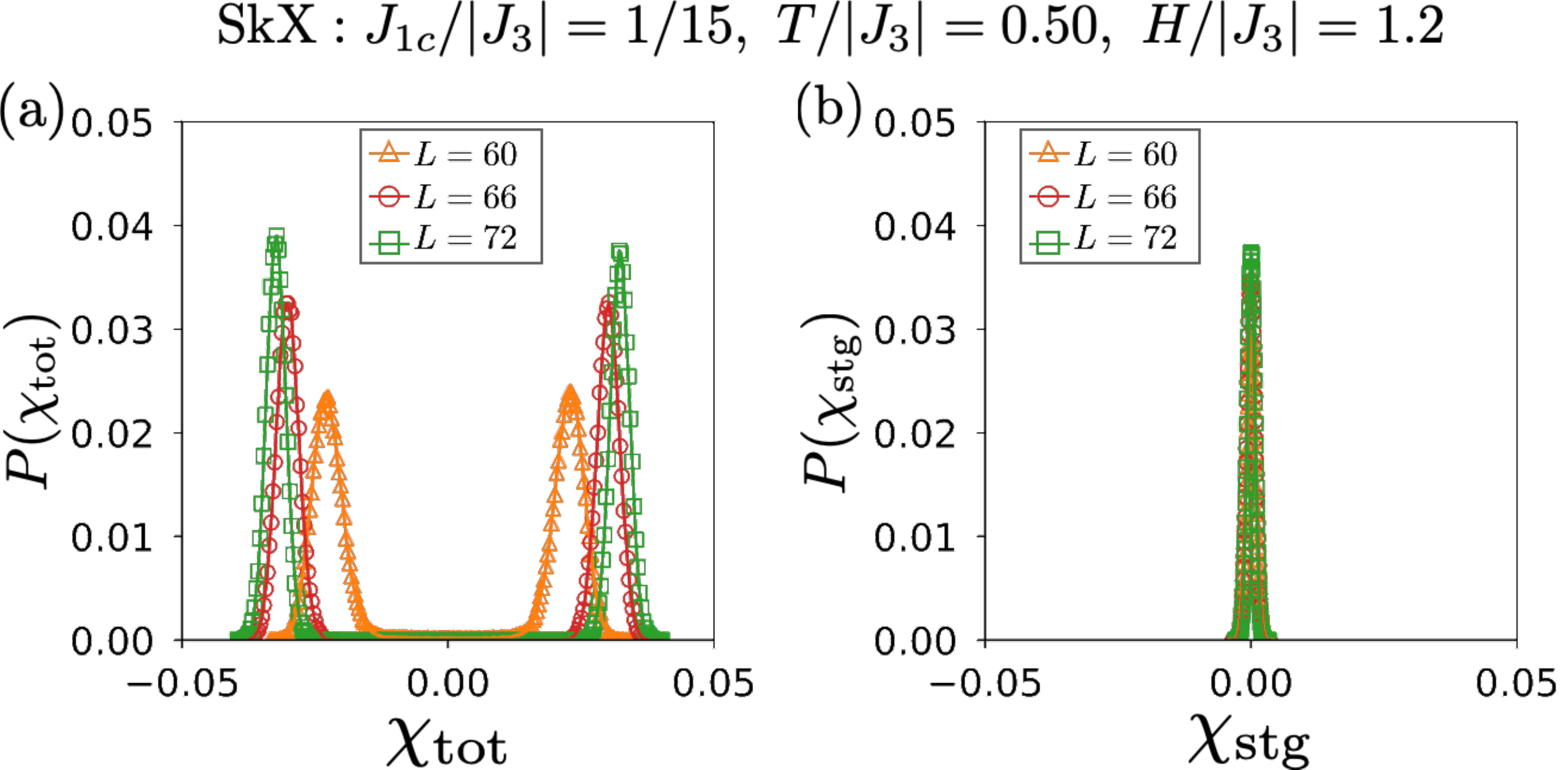}	
		\end{minipage}
	\end{tabular}
	\caption{
 The distribution of (a) the total scalar chirality $P(\chi_{{\rm tot}})$, and of (b) the staggered scalar chirality $P(\chi_{{\rm stg}})$, computed by the $T$-exchange run in the SkX state at $T/|J_3|=0.50$ and $H/|J_3|=1.2$  of the $J_1-J_3-J_{1c}$ model with the ferromagnetic interplanar coupling of $J_{1c}/|J_3|=1/15$. The lattice sizes are $L\times L\times \frac{2}{3}L$ with $L=60$, 66 and 72.
	}
	\label{3DFPD}
\end{figure}

\section{Antiferromagnetic interplanar coupling}

  In this section, we wish to deal with the case of the $J_1-J_3-J_{1c}$ 3D Heisenberg model on a stacked-triangular lattice with the antiferromagnetic nearest-neighbor interplanar coupling $J_{1c}<0$. Even in the case of the antiferromagnetic $J_{1c}$, the zero-field ground state is also a single-$q$ conical-spiral state in a wide parameter range of $0\leq J_1/|J_3| < 4$, characterized by the ordering wavevector ${\bm q}^*=({\bm q}_{xy}^*, \pi)$ with the same ${\bm q}_{xy}^*=(q_x^*,q_y^*)$ as that of the ferromagnetic-$J_{1c}$ model, where the single-$q$ conical-spiral in the triangular layer is stacked along the stacking ($z$) direction with alternating signs, i.e., $q_z^*=\pi$. Under applied magnetic fields, however, the antiferromagnetic $J_{1c}$ and applied fields compete with each other, and the situation becomes distinct from that of the ferromagnetic-$J_{1c}$ model.  

 Concerning the strength of the intraplanar couplings, we set $J_1/J_3=-1/3$ with $J_1>0$ and $J_3<0$ as in the previous section III and in Ref. \cite{OkuboChungKawamura}. Concerning the strength of $J_{1c}$, we study the two cases, i.e., moderately weak $J_{1c}/|J_3|=-\frac{1}{15}$, and even weaker $J_{1c}/|J_3|=-\frac{1}{60}$, as these two cases turn out to represent two distinct typical ordering behaviors of the model. The lattice sizes studied are $N=L\times L\times L_z$, 
with $L_z=\frac{1}{3}L$ ($r=\frac{1}{3}$) for both $J_{1c}/|J_3|=-\frac{1}{15}$ and $J_{1c}/|J_3|=-\frac{1}{60}$. $L$ is varied in the range $36 \leq L\leq 60$  for $J_{1c}/|J_3|=-\frac{1}{15}$, and in the range $60 \leq L\leq 90$  for $J_{1c}/|J_3|=-\frac{1}{60}$, with fixing the aspect ratio $r$. As the bulk physical quantities are expected not to depend on the sample shape, e.g., the aspect ratio, we set $r=\frac{1}{3}$ here in contrast to $r=\frac{2}{3}$ employed in the ferromagnetic $J_{1c}$ in the previous section, in order to save the computational cost.

\subsubsection{$J_{1c}/|J_3|=-\frac{1}{15}$}

 In this subsection, we consider the moderately weak antiferromagnetic nearest-neighbor interplanar coupling of $J_{1c}/|J_3|=-\frac{1}{15}$. Since the SkX state stabilized at finite fields has a net magnetization along the field which competes with the antiferromagnetic interplanar coupling, the SkX state for the antiferromagnetic $J_{1c}$ becomes less stable. Indeed, even for the moderately weak antiferromagnetic interplanar coupling of $J_{1c}/|J_3|=-\frac{1}{15}$, the SkX state turns out to be gone entirely from the $T$-$H$ phase diagram. The only ordered state is a single-$q$ conical-spiral state with a spontaneously broken $C_3$ symmetry, with the ordering wavevector ${\bm q}^*=({\bm q}_{xy}^*, \pi)$. The obtained $T$-$H$ phase diagram is shown in Fig. 8. No multiple-$q$ state is realized. 

 We add that we also repeat the calculation for a different value of the aspect ratio of $r=\frac{2}{3}$, the value employed in the ferromagnetic-$J_{1c}$ case in the previous section, to observe that the phase diagram is essentially the same, just as expected.

\begin{figure}[t]
	\centering
	\begin{tabular}{c}
		\begin{minipage}{\hsize}
			\includegraphics[width=\hsize]{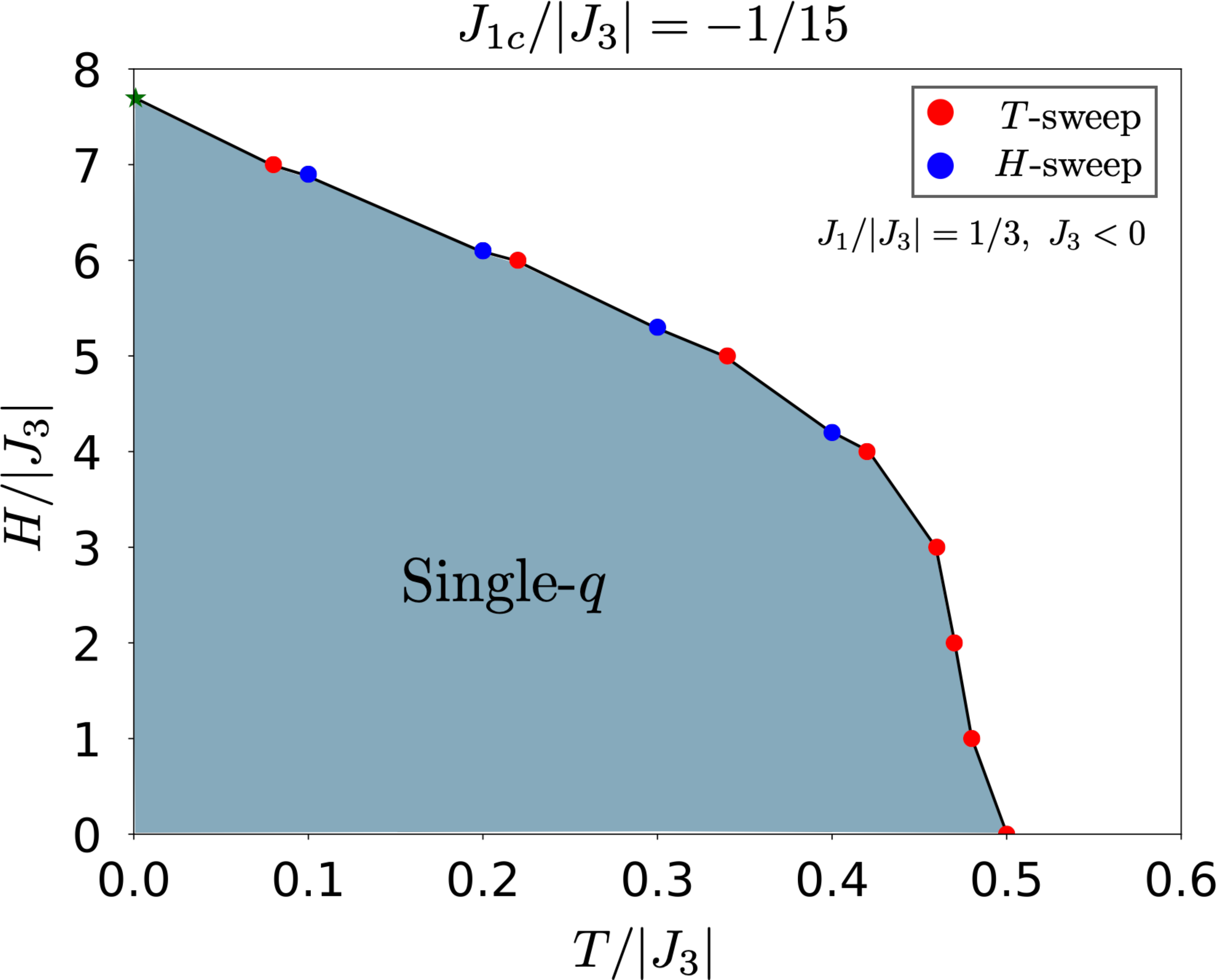}	
		\end{minipage}
	\end{tabular}
	\caption{
The temperature ($T$) versus magnetic field ($H$) phase diagram of the frustrated $J_1-J_3-J_{1c}$ classical Heisenberg model on a 3D stacked-triangular lattice with the antiferromagnetic nearest-neighbor interplanar coupling of $J_{1c}/|J_3|=-1/15$ as determined by MC simulations, where $J_1$ and $J_3$ are the ferromagnetic nearest-neighbor and the antiferromagnetic third-neighbor intraplanar couplings with $J_1/|J_3|=1/3$. Phase boundaries are determined both by $T$- and $H$-sweeps.
	}
	\label{3DFPD}
\end{figure}

\subsubsection{$J_{1c}/|J_3|=-\frac{1}{60}$}

 In search for the possible nontrivial $T$-$H$ phase diagram containing the SkX state for the antiferromagnetic $J_{1c}$, we examine the case of even weaker $J_{1c}$. In this subsection, we consider the interplanar coupling of $J_{1c}/|J_3|=-\frac{1}{60}$.  In fact, if $J_{1c}$ is taken to be this small value, the SkX revives at intermediate fields and at finite temperatures, together with the $Z$ phase and the double-$q$ phase.
\begin{figure}[t]
	\centering
	\begin{tabular}{c}
		\begin{minipage}{\hsize}
			\includegraphics[width=\hsize]{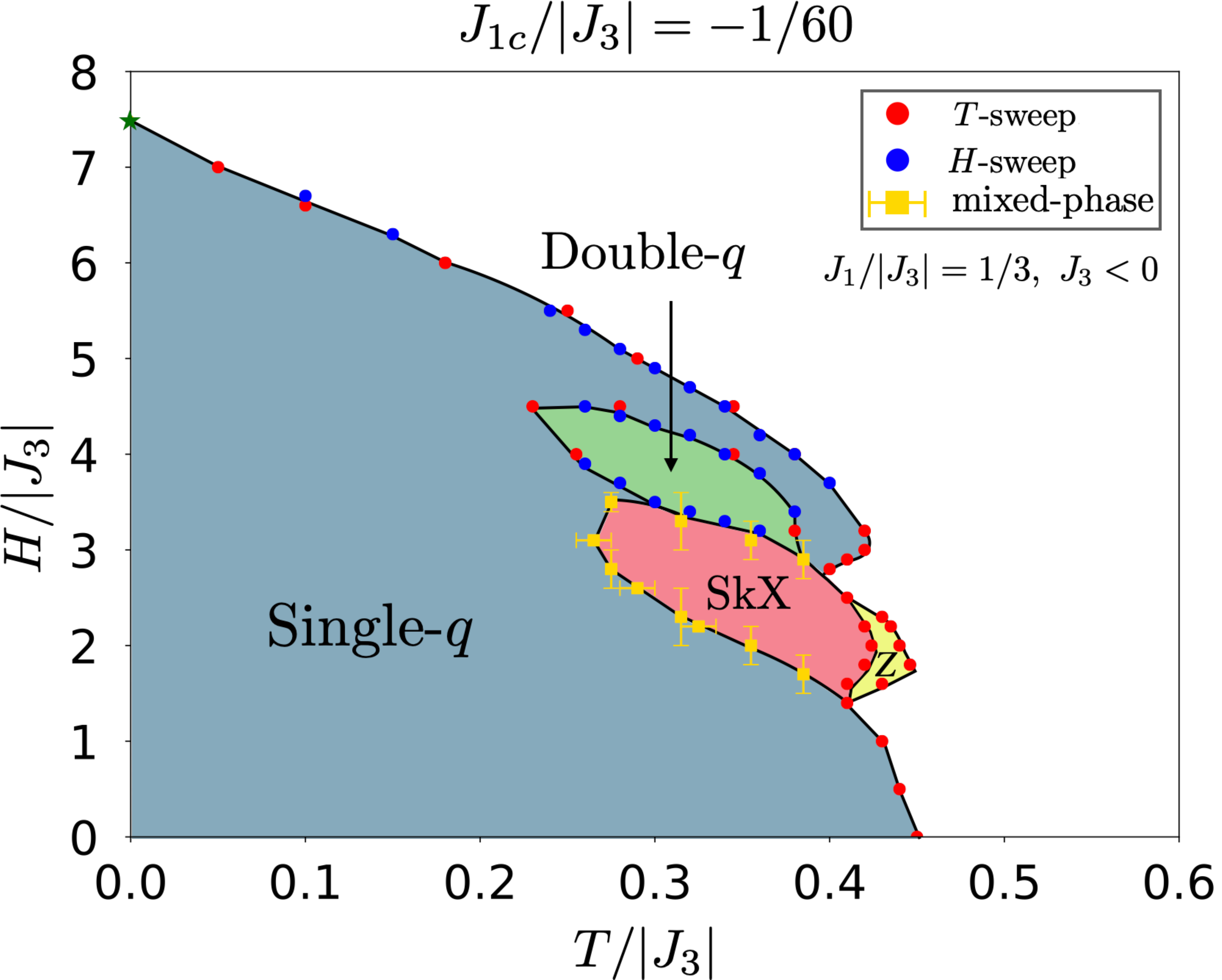}	
		\end{minipage}
	\end{tabular}
	\caption{
The temperature ($T$) versus magnetic field ($H$) phase diagram of the frustrated $J_1-J_3-J_{1c}$ classical Heisenberg model on a 3D stacked-triangular lattice with the weaker antiferromagnetic nearest-neighbor interplanar coupling of $J_{1c}/|J_3|=-1/60$ as determined by MC simulations, where $J_1$ and $J_3$ are the ferromagnetic nearest-neighbor and the antiferromagnetic third-neighbor intraplanar couplings with $J_1/|J_3|=1/3$. Phase boundaries are determined both by $T$- and $H$-sweeps, the mixed-phase method \cite{mixed phase} also employed.  
	}
	\label{3DFPD}
\end{figure}

 The computed $T$-$H$ phase diagram is shown in Fig. 9. Although the main features of the phase diagram look more or less similar to the corresponding phase diagrams of the 2D model \cite{OkuboChungKawamura} and of the 3D model with the ferromagnetic $J_{1c}$ shown in Fig. 1, the stability region of the multiple-$q$ phases is considerably reduced relative to the single-$q$ phase.

 In Fig. 10, we show the $T$-dependence of several physical quantities at a field $H/|J_3|=2.2$ in the $T$ range including the paramagnetic, $Z$ and SkX phases. In contrast to the ferromagnetic-$J_{1c}$ case, $T$-annealing runs turn out to fail to fully thermalize the SkX state of the antiferromagnetic-$J_{1c}$ model, especially, fail to reproduce the correct staking patterns of SkX layers in thermal equilibrium (details are given below). In order to fully thermalize the SkX phase, we need to employ the $T$-exchange runs, which are certainly possible for the sizes $L\leq 72$, but unfortunately turn out to fail for the larger sizes $L\geq 78$. Hence, in Fig. 10, the data for the sizes $L\leq 72$ are taken by the fully equilibrated $T$-exchange runs which cover the $T$ range down to the SkX phase, while the data for the sizes $L\geq 78$ are taken by the $T$-annealing runs which are limited to relatively high-$T$ range down to the $Z$ phase. In the $Z$ phase, by contrast, thermalization is easier, and even the $T$-annealing runs yield fully thermalized results for the largest size $L=90$.


 The $T$-dependence of the specific heat is shown in Fig. 10(a). As can be seen from the figure, the specific heat in this $T$ range exhibits a change of behavior from smaller sizes of $L\lesssim 72$ to larger sizes of $L\gtrsim 72$, i.e., a single-peak structure observed for smaller sizes $L\lesssim 72$ changes into a double-peak structure for larger size $L\gtrsim 72$ by developping a dull peak (kink) at a higher temperature $T/|J_3| \simeq 0.44$. The occurrence of a weak double-peak anomalies for larger sizes is consistent with the occurrence of the $Z$ phase in the $T$ range of $0.42\lesssim T/|J_3|\lesssim 0.44$.

 The $T$-dependence of the total scalar chirality $\overline \chi_{{\rm tot}}$ is shown in Fig. 10(b). At lower-$T$ range corresponding to the SkX state,  with increasing the system size $L$, $\overline \chi_{{\rm tot}}$ tends to increase monotonically tending to a nonzero value, consistently with the existence of the SkX phase. At somewhat higher-$T$ range $0.42\lesssim T/|J_3|\lesssim 0.44$, by contrast, a changeover similar to the one observed in the specific heat $C$ is observed in the size dependence of $\overline \chi_{{\rm tot}}$. Namely, although $\overline \chi_{{\rm tot}}$ tends to grow with increasing $L$ for $L\lesssim 72$, it tends to be suppressed for $L\gtrsim 72$ showing a size crossover as can be seen from the inset of Fig. 10(b), suggesting that $\overline \chi_{{\rm tot}}$  in this $T$ range eventually vanishes for sufficiently large $L$. This observation is also consistent with the existence of the $Z$ phase with a vanishing  $\overline \chi_{{\rm tot}}$ in the $L\rightarrow \infty$ limit.
\begin{figure}[t]
	\centering
	\begin{tabular}{c}
		\begin{minipage}{\hsize}
			\includegraphics[width=\hsize]{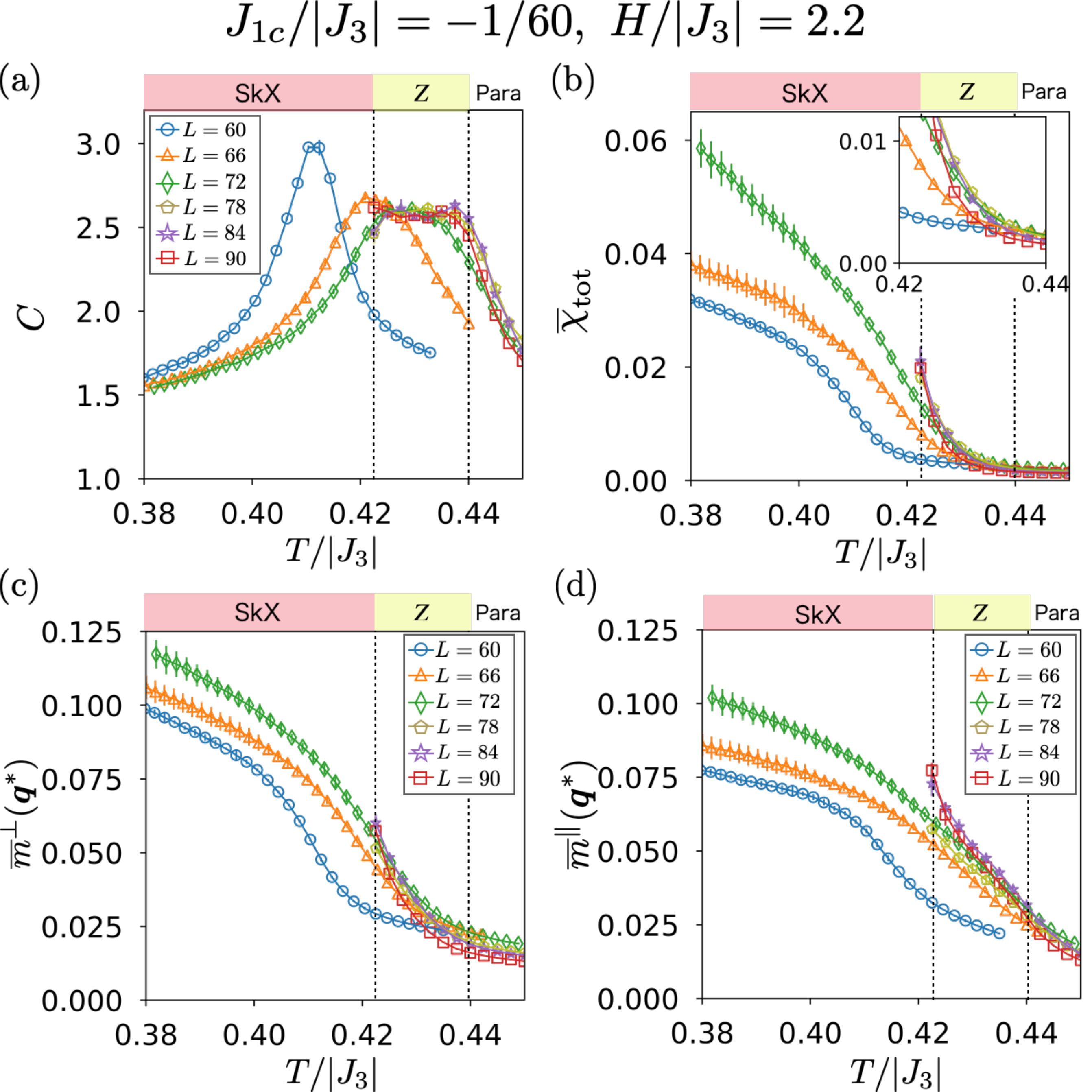}	
		\end{minipage}
	\end{tabular}
	\caption{
The temperature and size dependence of (a) the specific heat, (b) the total scalar chirality, (c) the transverse ($\perp$) and (d) the  longitudinal  ($\parallel$) components of the magnetic order parameters associated with the ordering wavevectors ${\bm q}^*=(q_x^*,q_y^*,\pi)$ of the frustrated $J_1-J_3-J_{1c}$ Heisenberg model on a stacked-triangular lattice with the weaker antiferromagnetic nearest-neighbor interplanar coupling of $J_{1c}/|J_3|=-1/60$. The lattice sizes are $L\times L\times \frac{1}{3}L$ with $L=60$, 66, 72, 78, 84 and 90. The data for the sizes $L\leq 72$ are taken by the $T$-exchange runs, while the data for the sizes $L\geq 78$ are taken by the $T$-annealing runs. 
	}
	\label{3DFPD}
\end{figure}

 In Fig. 10(c) and 10(d), we show the $T$-dependence of the magnetic order parameters associated with the ordering wavevector ${\bm q}^*=({\bm q}_{xy}^*,\pi)$ for each case of (c) the spin-transverse ($S_x,S_y$) components $\overline m^\perp ({\bm q}^*)$, and of (d) the spin-longitudinal $S_z$ component $\overline m^\parallel ({\bm q}^*)$, where we set $q_z^*=\pi$ in view of the antiferromagnetic character of the interplanar coupling $J_{1c}$. (As will be shown below, the Bragg component actually appears also at $q_z=0$ in the SkX state.) In the lower-$T$ range corresponding to the SkX phase, both $\overline m^\perp ({\bm q}^*)$ and $\overline m^\parallel ({\bm q}^*)$ grow tending to a nonzero value. In the intermediate $T$ range, there again occurs a size crossover between smaller sizes of $L\lesssim 72$ and larger sizes of $L\gtrsim 72$. Namely, in the $T$ range of $0.42\lesssim T/|J_3|\lesssim 0.44$, while both $\overline m^\perp ({\bm q}^*)$ and $\overline m^\parallel ({\bm q}^*)$ grow with increasing $L$ for $L\lesssim 72$, $\overline m^\perp ({\bm q}^*)$ tends to be suppressed for $L\gtrsim 72$ while $\overline m^\parallel ({\bm q}^*)$ continues to grow. These data are suggestive of the stabilization of the $Z$ phase in this $T$ range, characterized by nonzero $\overline m^\parallel ({\bm q}^*)$ but vanishing $\overline m^\perp ({\bm q}^*)$, consistently with the observations from $C$ and $\overline \chi_{{\rm tot}}$.   

 Typical spin structure factors $S(\bm{q})$ of the SkX state are shown in the $(q_x,q_y)$ plane in Fig. 11, i.e., $S^\perp({\bm q})$ and $S^\parallel ({\bm q})$ for $q_z=0$ in (a) and (b), and those for $q_z=\pi$ in (c) and (d), respectively. While the SkX state preserves the lattice $C_3$ symmetry associated with the three ordering wavevectors ${\bm q}_1^*$, ${\bm q}_2^*$ and ${\bm q}_3^*$, the Bragg peaks appear not only at $q_z=\pi$ as favored by the antiferromagnetic $J_{1c}$, but also at $q_z=0$ favored by the magnetic field, in contrast to the ferromagnetic $J_{1c}$ where the Bragg peaks appear only at $q_z=0$. 
 This observation suggests that the SkX layers stack in somewhat alternating way, and there occurs some relative sliding between the arrangements of the two SkX's on the adjacent triangular layers, as schematically illustrated in Fig. 5(b). 

\begin{figure}[t]
	\centering
	\begin{tabular}{c}
		\begin{minipage}{\hsize}
			\includegraphics[width=\hsize]{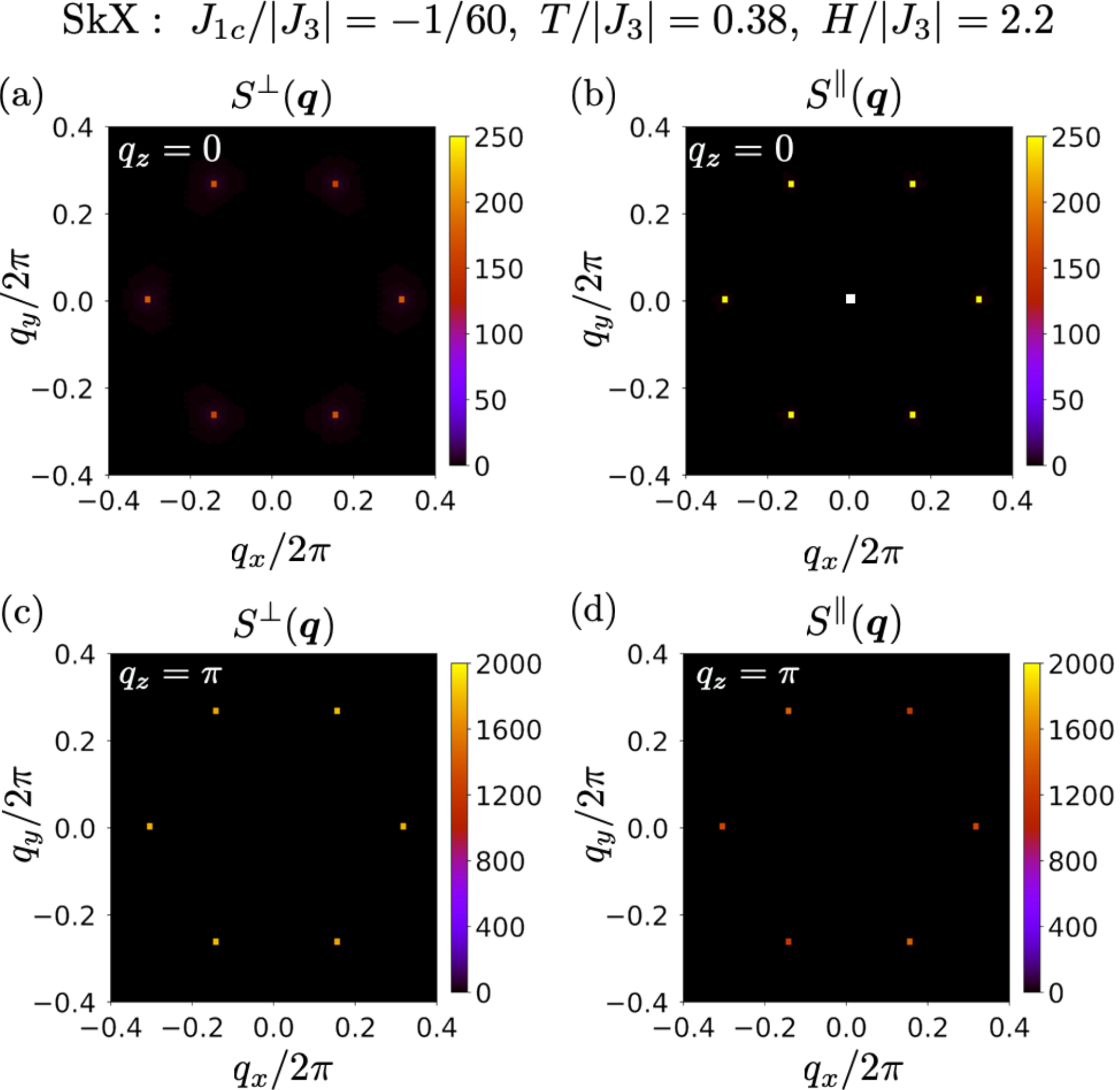}	
		\end{minipage}
	\end{tabular}
	\caption{
Spin structure factors in the SkX state at $T/|J_3|=0.38$ and $H/|J_3|=2.2$  for the weaker antiferromagnetic nearest-neighbor interplanar coupling of $J_{1c}/|J_3|=-1/60$ are shown in the $(q_x,q_y)$ plane,  (a,b) at $q_z=0$, and (c,d) at $q_z=\pi$, representing (a,c) the perpendicular $S^\perp (\bm{q})$, and (b,d) the parallel $S^\parallel (\bm{q})$. The lattice size is $72\times 72\times 24$ ($L=72$). The state is generated by the $T$-exchange run, while the $T$-exchange process is cut off during the measurements.
}
\end{figure}

 More direct information about the manner of the SkX-layer stacking might be obtained by examining the real-space spin configurations in the SkX state. Typical real-space spin configurations of the SkX state obtained from the fully equilibrated $T$-exchange simulation are shown in Fig. 12 for three successive triangular layers. 
 One can see from the figure that the skyrmion core now forms, not a direct on-top stack as in the ferromagnetic-$J_{1c}$ case, but rather $ABABAB\cdots $-type stack where the skyrmion core of the next layer, say, layer 2, is located at the center position of the triangle formed by the skyrmion cores in the original layer, say, layer 1. Such a slided stacking arises due to the competition between the uniform stacking favored by the magnetic field and the antiparallel spin alignment favored by the antiferromagnetic $J_{1c}$. Similar $ABABAB\cdots $-type stacking pattern of SkX layers was also reported for the 3D frustrated Heisenberg model on a stacked-triangular lattice with moderately strong easy-axis anisotropy \cite{LinBatista3D}.

 We note that such $ABABAB\cdots $-type stacking is reproducible in the $T$-exchange runs, and thereby is expected to be a truly stable stacking pattern in thermal equilibrium. However, if one simply anneals or quenches the system from high $T$ to the lower-$T$ SkX state without paying attention to equilibration, other types of stacking patterns including apparently random stacking patterns often appear as metastable states, even including the random stacking of both SkX and anti-SkX layers. This point will be further discussed below in this section and in \S V.

\begin{figure}[t]
	\centering
	\begin{tabular}{c}
		\begin{minipage}{\hsize}
			\includegraphics[width=\hsize]{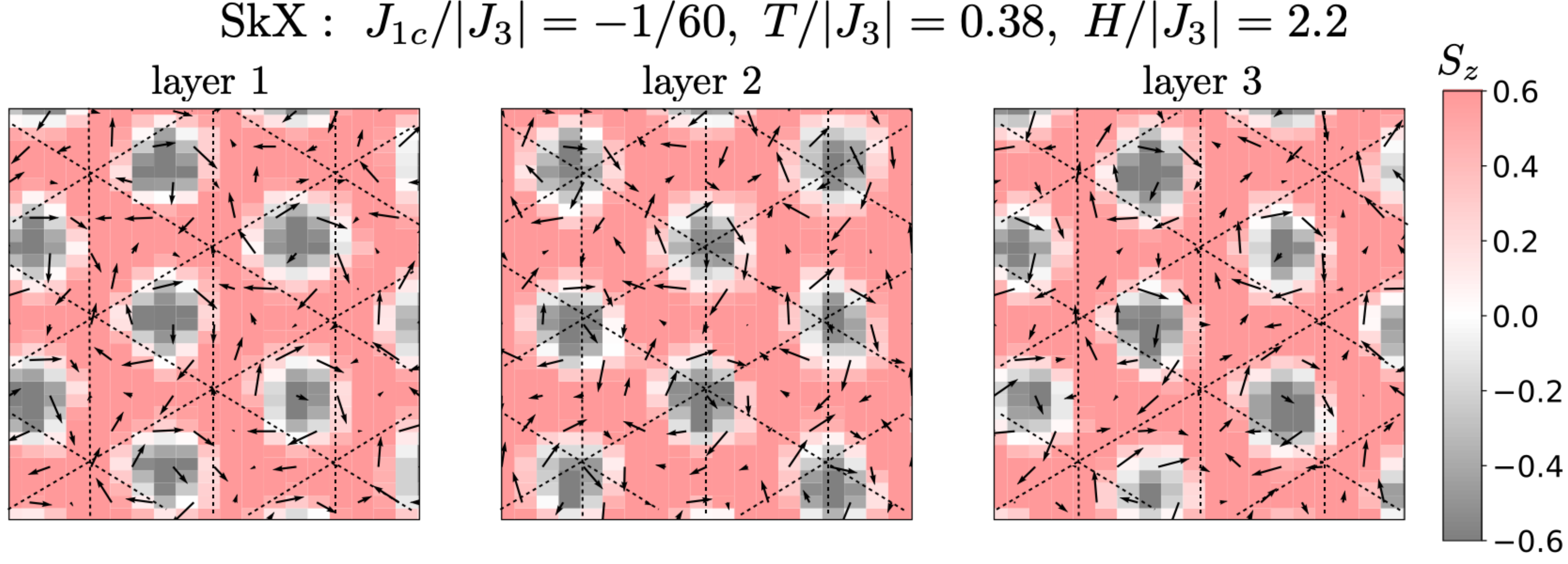}	
		\end{minipage}
	\end{tabular}
	\caption{
Typical real-space spin configurations in the SkX state at  $T/|J_3|=0.38$ and $H/|J_3|=2.2$ for the weaker antiferromagnetic nearest-neighbor interplanar coupling of $J_{1c}/|J_3|=-1/60$ for three successive triangular layers, from layer 1 to layer 3. The color represents the spin $S_z$ component, while the arrow represents the direction of the spin-transverse ($S_x,S_y$) components. To reduce the thermal noise, short-time averaging of 50 MCS is made. The figure represents a part of the $72\times 72\times 24$ ($L=72$) lattice with a common $xy$ section among layers 1-3.  The state is generated by the $T$-exchange run, while the $T$-exchange process is cut off during the short-time averaging.
}
	\label{3DFPD}
\end{figure}

 Due to the underlying $Z_2$ spin-mirror symmetry, both the SkX and the anti-SkX states are equally possible in the SkX phase of the present isotropic model. This is demonstrated in Figs. 13(a) and (b) where the color plots of typical real-space local scalar chirality configurations of (a) the SkX and (b) the anti-SkX states are shown for three successive triangular layers for the SkX phase. One can see that in the SkX phase the scalar chiralities are uniformly ordered either to negative (SkX state) or positive (anti-SkX state) value just as in the ferromagnetic-$J_{1c}$ case. 

 We also compute the Fourier transform of the layer chirality, $S_\chi (q_z)$, defined by
\begin{equation}
S_\chi (q_z) = \frac{1}{L_z} \left\langle \left| \sum_{i_z=1}^{L_z} \chi_{{\rm layer}} (i_z) e^{-i q_z\cdot i_z} \right|^2 \right\rangle, 
\end{equation}
where $\chi_{{\rm layer}}(i_z)$ is the total scalar chirality of the $i_z$-th triangular layer of the stacked-triangular lattice defined by
\begin{equation}
\chi_{{\rm layer}} (i_z) = \frac{1}{2L^2} \left( \sum_{\bigtriangleup\in {\rm layer}:i_z} \chi_\bigtriangleup + \sum_{\bigtriangledown\in {\rm layer}:i_z} \chi_\bigtriangledown \right).
\end{equation}

 The $q_z$-dependence of $S_\chi (q_z)$ computed by equilibrated $T$-exchange simulations is shown in Fig. 13(c). As can be seen from the figure,  with increasing $L$, the uniform $q_z=0$ component continues to grow, consistently with the divergent behavior. This indicates that the SkX staking along the $z$-direction in the SkX phase, when viewed via the scalar-chirality degrees of freedom, is indeed uniform ($q_z=0$) in spite of its slided stacking pattern when viewed via the spin degrees of freedom. 
\begin{figure}[t]
	\centering
	\begin{tabular}{c}
		\begin{minipage}{\hsize}
			\includegraphics[width=\hsize]{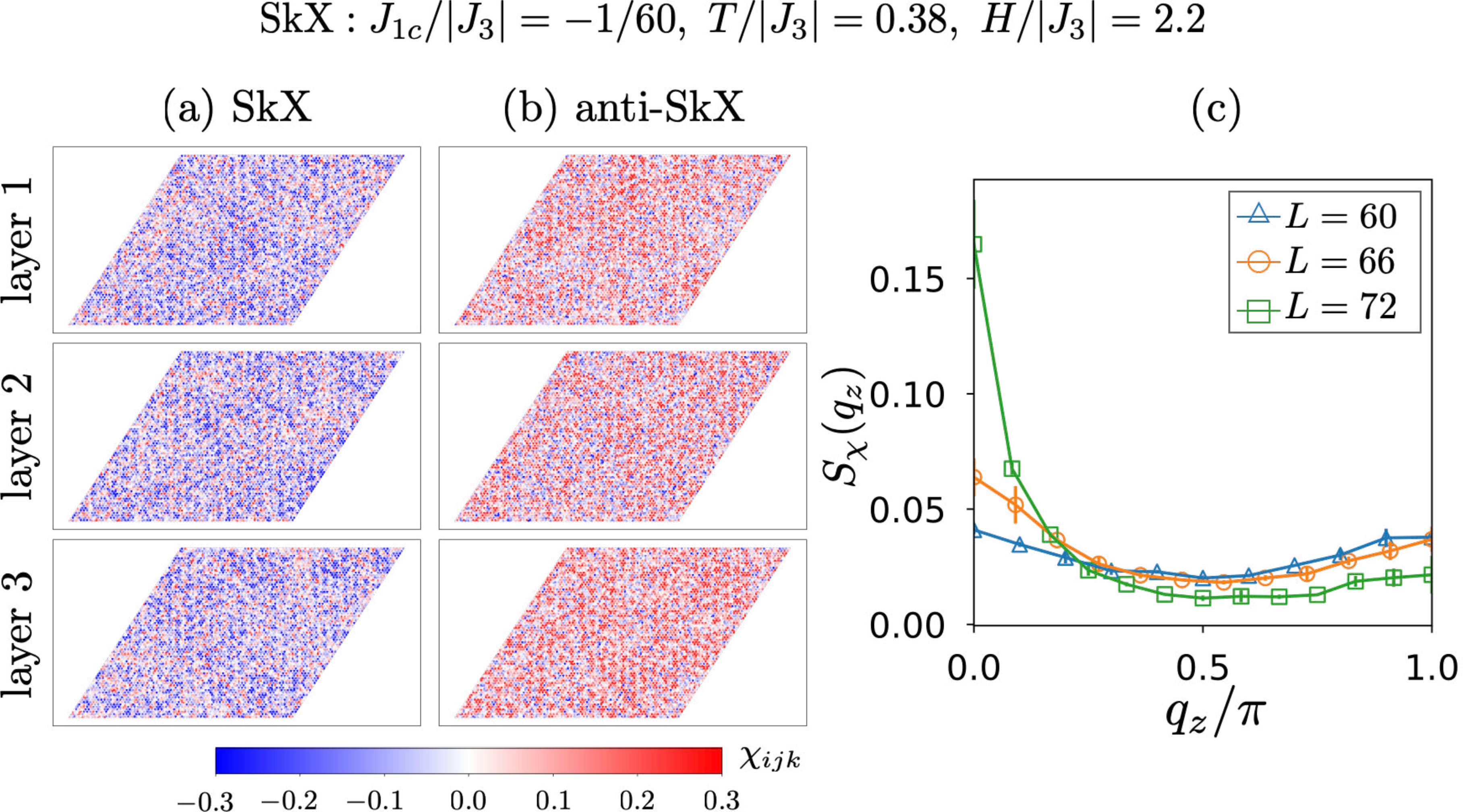}	
		\end{minipage}
	\end{tabular}
	\caption{
 Color plots of typical real-space local scalar chirality configurations of (a) the SkX state and (b) the anti-SkX state for three successive triangular layers, from layer 1 to layer 3, and (c) the $q_z$-dependence of the Fourier-transformed layer chirality $S_\chi(q_z)$, in the SkX phase at $T/|J_3|=0.38$ and $H/|J_3|=2.2$ for the weaker antiferromagnetic nearest-neighbor interplanar coupling of $J_{1c}/|J_3|=-1/60$. In (a) and (b), to reduce the thermal noise, short-time averaging of 50 MCS is made. The figure represents a $72\times 72$ triangular sheet of the 3D stacked-triangular lattice of the size $72\times 72\times 24$. In (c), the lattice sizes are $L\times L\times \frac{1}{3}L$ with $L=60, 66$ and 72.  In (a, b), the state is generated by the $T$-exchange run, while the $T$-exchange process is cut off during the short-time averaging. In (c), the data are taken by the $T$-exchange runs.
	}
	\label{3DFPD}
\end{figure}

 In Figs. 14(a) and (b), we show the $q_z$-dependence of the perpendicular and the parallel spin structure factors $S^\perp (q_x^*,q_y^*,q_z)$ and $S^\parallel (q_x^*,q_y^*,q_z)$, respectively. In both $S^\perp (q_x^*,q_y^*,q_z)$ and $S^\parallel (q_x^*,q_y^*,q_z)$, while the $q_z=\pi$ component dominates over the $q_z=0$ component in their intensities, both the $q_z=\pi$ and the $q_z=0$ components exhibit a divergent-like $L$-dependence, suggesting the occurrence of the Bragg peaks both at $q_z=\pi$ and $q_z=0$. 
 Then, in the full ${\bm q}$-space, there exist six independent Bragg peaks at (${\bm q}_{1,xy}^*,0$),  (${\bm q}_{2,xy}^*,0$), (${\bm q}_{3,xy}^*,0$), (${\bm q}_{1,xy}^*,\pi$), (${\bm q}_{2,xy}^*,\pi$) and (${\bm q}_{3,xy}^*,\pi$), in contrast to only three independent Bragg peaks for the case of ferromagnetic $J_{1c}$.
\begin{figure}[t]
	\centering
	\begin{tabular}{c}
		\begin{minipage}{\hsize}
			\includegraphics[width=\hsize]{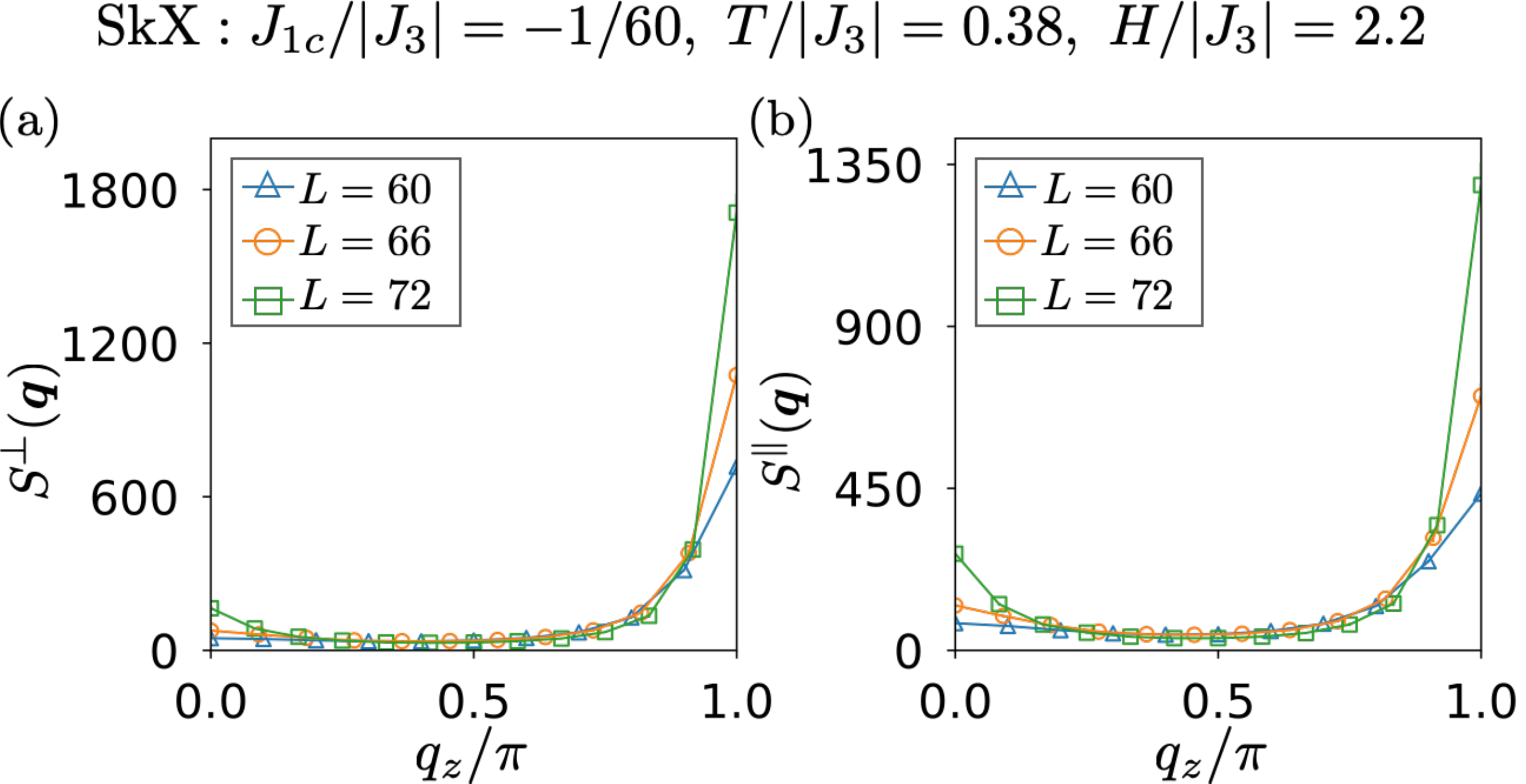}	
		\end{minipage}
	\end{tabular}
	\caption{
The $q_z$-dependence of the spin structure factors of (a) the spin-transverse components $S^\perp({\bm q})$, and (b) the spin-longitudinal component $S^\parallel({\bm q})$, where $(q_x,q_y)$ is set to $(q_x^*, q_y^*)$, in the SkX state at $T/|J_3|=0.38$ and $H/|J_3|=2.2$ for the weaker antiferromagnetic nearest-neighbor interplanar coupling of $J_{1c}/|J_3|=-1/60$.  The lattice sizes are $L\times L\times \frac{1}{3}L$ with $L=60, 66$ and 72. The data are taken by the $T$-exchange runs.
	}
	\label{3DFPD}
\end{figure}

\vskip 0.6cm
\noindent
{\it Reference calculation of the SkX-layers stacking\/}
\vskip 0.2cm

 In order to get some more insight into the observed SkX stacking pattern, we perform an analytical reference calculation. First, we determine the most energetically favorable stacking pattern for just two adjacent SkX layers 1 and 2. Since the spin configuration of the 2D SkX state in each layer observed by MC is more or less similar to the one observed in the corresponding 2D $J_1-J_3$ model, we take a mean-field spin configuration for the 2D $J_1-J_3$ model which has turned out to well describe the SkX state of the 2D model \cite{OkuboChungKawamura}, given by 
\begin{eqnarray}
 {\bm S}_{i,xy} &=& I_{xy}\sum_{j} \sin({\bm q}^\ast_j\cdot{\bm r}_i+\theta_j) {\bm e}_j, \nonumber
\\
 S_{i,z} &=& I_z\sum_{j} \cos({\bm q}^\ast_j\cdot {\bm r}_i+\theta_j) + m_z, 
\label{SkX}
\end{eqnarray}
where the $j$ sum is taken over the three ${\bm q}_j^*$ modes $j=1,2,3$, $I_{xy}$ and $I_z$ are $T$-dependent constants, $m_z$ is a uniform magnetization induced by an external field, $\theta_j$ ($j=1,2,3$) are phase factors satisfying the condition $\cos(\theta_1+\theta_2+\theta_3)=-1$, and $\bm{e}_j$ ($j=1,2,3$) are arbitrary three unit vectors lying in the spin-transverse $(S_x, S_y)$ plane satisfying $\sum_j\bm{e}_j=\bm{0}$. The remaining two degrees of freedom associated with $\theta_j$ ($j=1,2, 3$) correspond to the translation degrees of freedom of the SkX against the original spin triangular lattice. The phases $\theta_j^{(1)}$ of the first layer $1$ can be taken to be $\theta_1^{(1)}=\theta_2^{(1)}=\theta_3^{(1)}=\pi/3$, without loosing generality.

 First, we try to simulate the spin orientations of the single SkX-layer obtained by our present MC simulation at $T/|J_3|=0.38$ and $H/|J_3|=2.2$ by Eqs. (15), to find that $I_{xy}=I_z=0.24$ and $m_z=0.3$ can well simulate the MC result. The phases of the next layer 2, $\theta_j^{(2)}$, minimizing the antiferromagnetic interlayer energy is then searched for, where the spin length is rescaled to unity with keeping its orientation given by Eqs. (15),  which yields $\theta_1^{(2)}=\pi$, $\theta_2^{(2)}=-\pi$, $\theta_3^{(2)}=\pi$. The layer sliding described by these phase values just corresponds to the neighboring-layer stacking observed in our MC for the 3D model, i.e., the skyrmion core of the layer 2 is located at the center position of the triangle formed by the skyrmion cores of the layer 1. Thus, the sliding pattern of the two adjacent SkX layers observed by our MC can be understood from a simple energy consideration. The resulting real-space spin configurations of the layers 1 and 2 are shown in Fig. 15(a). 

 While the underlying $Z_2$ symmetry dictates that the SkX and the anti-SkX have equal energies for a single layer, for the present two-layers system, the SkX-SkX and the SkX - anti-SkX configurations generally have different energies. Thus, we also perform a similar energy optimization calculation for the SkX - anti-SkX configurations, to observe that the minimum energy is obtained for the present parameter choice when the anti-skyrmion core of the layer 2 is located at the midpoint of the edge of the triangle formed by the skyrmion cores of the layer 1, and that the optimized energy is slightly higher than that of the SkX-SkX configuration, by about $\sim 0.1$\%. The result seems consistent with our MC observation that, in the SkX phase, each layer exhibits the topological charge, or the layer scalar chirality, of the same sign. Hence, the stable stacking pattern in the SkX phase is suggested to be the stacking of only SkX (or only anti-SkX) layers.
%
%
\begin{figure}[t]
	\centering
	\begin{tabular}{c}
		\begin{minipage}{\hsize}
			\includegraphics[width=\hsize]{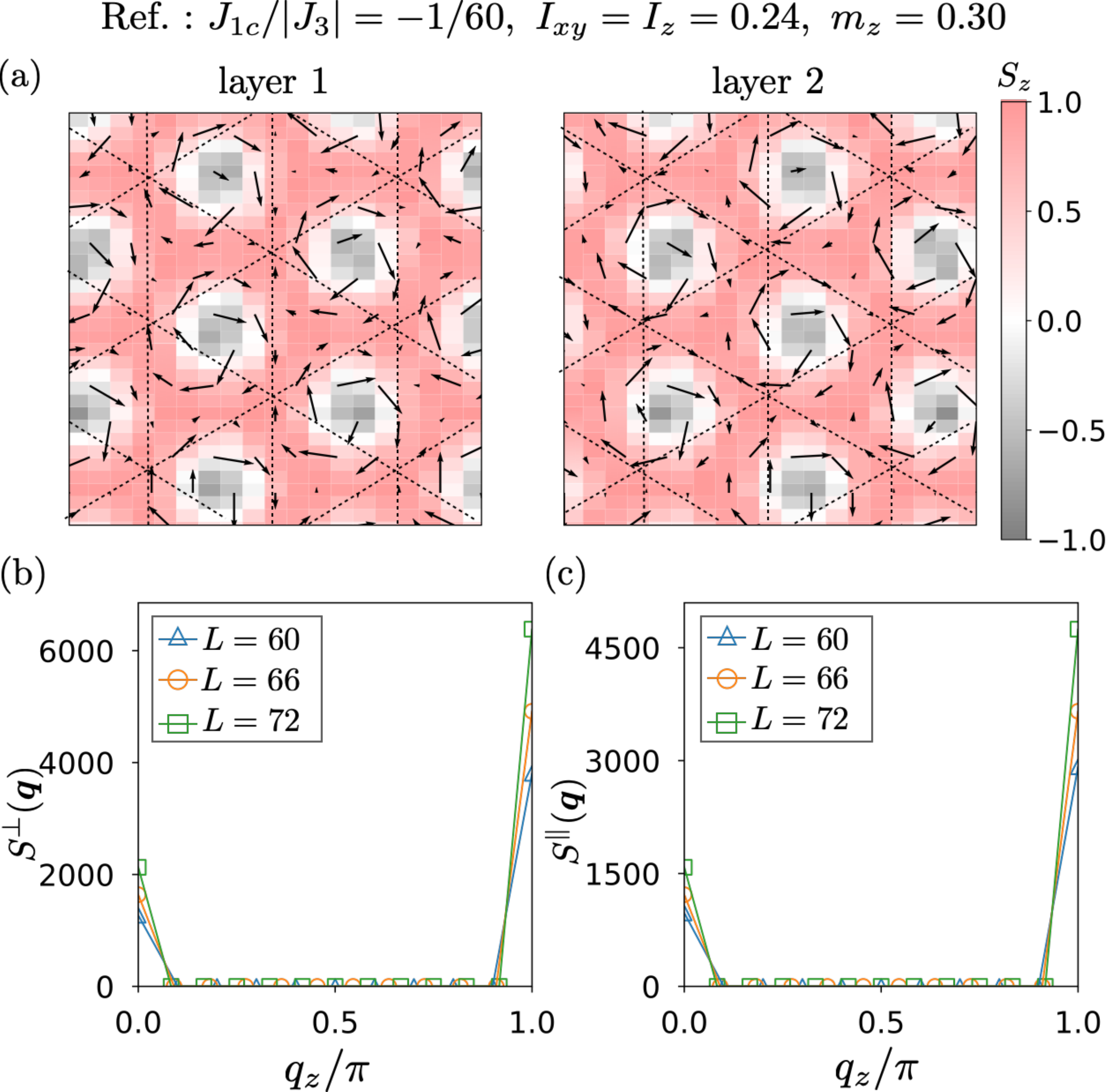}	
		\end{minipage}
	\end{tabular}
	\caption{
(a) Real-space spin configurations, and the $q_z$-dependence of the spin structure factors of (b) the spin-transverse components $S^\perp({\bm q})$ and (c) the spin-longitudinal component $S^\parallel({\bm q})$ where $(q_x,q_y)$ is set to $(q_x^*, q_y^*)$, obtained by the mean-field-type reference calculation on the two-layers model (see the text for details). The mean-field parameters are taken to be $I_{xy}=I_z=0.24$ and $m_z=0.30$ to simulate the SkX state observed by MC. In (b, c), the lattice sizes are $L\times L\times \frac{1}{3}L$ with $L=60, 66$ and 72.
	}
	\label{3DFPD}
\end{figure}

 From symmetry, the center position of the triangle associated with the skyrmion superlattice of the layer 1,  $A$, is twofold degenerate in the next layer 2, i.e., $B$ or $C$, so that the stackings $AB$ and $AC$ are equally possible. Such degeneracy could lead to energetically degenerate interlayer SkX stacking patterns in the bulk 3D SkX state, not just $ABABAB\cdots$, but also $ABCABC\cdots $, $ABACBC\cdots$, and infinitely many others. Our MC has indicated that, among such infinitely-many stacking patterns,  $ABABAB\cdots$-type stacking is chosen and stable. This selection is most probably due to the order-from-disorder effect \cite{Villain,KawamuraSW,Henley}. In the present context, the order-from-disorder mechanism suggests that the $ABABAB\cdots$-type stacking possesses the largest number of low-energy excited states just above it, i.e., possesses the highest entropy, among possible many other stacking states.
 
 Once we accept this and assume the $ABABAB\cdots$-type SkX stacking pattern, we can compute the Bragg intensity of the associated spin structure factors, $S^\perp({\bm q})$ and $S^\parallel({\bm q})$, on the basis of Eqs. (15) (without the spin-length rescaling here). We then find that the Bragg peaks indeed appear at six independent points in the $q$-space,  (${\bm q}_{1,xy}^*,0$),  (${\bm q}_{2,xy}^*,0$), (${\bm q}_{3,xy}^*,0$), (${\bm q}_{1,xy}^*,\pi$), (${\bm q}_{2,xy}^*,\pi$) and (${\bm q}_{3,xy}^*,\pi$). The Bragg intensities at $q_z=\pi$ and at $q_z=0$ are calculated as $S({\bm q}^*_{xy}, \pi) = \frac{3}{4}I_{xy}^2N$ and $S({\bm q}^*_{xy}, 0) = \frac{1}{4}I_{xy}^2N$, yielding their ratio 3:1 for both $S^\perp({\bm q})$ and $S^\parallel({\bm q})$ irrespective of the $I_{xy}$-value. The resulting $q_z$-dependence is shown in Figs. 15(b) and (c) for $S^\perp(q_{xy}^*,q_z)$ and $S^\parallel(q_{xy}^*,q_z)$, respectively. In our MC simulation, we indeed observe the six independent Bragg peaks at these ${\bm q}$-positions in the SkX state, although the intensity ratio observed by MC seems to be somewhat greater than 3:1.

 The reason of such a minor quantitative deviation is not entirely clear. While we have checked that the mean-field spin configuration given by Eqs. (15), originally derived for a single-layer SkX state, gives a reasonable description of the SkX of the present 3D model, it might still be affected somewhat by the antiferromagnetic interplanar coupling competing with the applied magnetic field, and after all, the mean-field approximation neglects the effect of fluctuations, {\it etc\/}. Yet, the approximation captures the essential qualitative features of the 3D SkX structure even for the antiferromagnetic $J_{1c}$.

\vskip 0.6cm

 Now, returning to our MC results, we wish to discuss the $Z$ phase. Typical spin structure factors $S(\bm{q})$ of the $Z$ state obtained by the $T$-annealing run for the largest size $L=90$ are shown in the $(q_x,q_y)$ plane in Fig. 16, i.e., $S^\perp(\bm{q})$ and $S^\parallel (\bm{q})$ for $q_z=0$ in (a) and (b), and the ones for $q_z=\pi$ in (c) and (d), respectively. As can be seen from these figures, the peaks of $S^\perp(\bm{q})$ are considerably broader than those of $S^\parallel (\bm{q})$, suggesting that the only longitudinal component exhibits a magnetic long-range order while the transverse component exhibits a short-range order only. This observation is consistent with the behavior of the magnetic order parameters shown in Fig. 10(c) and 10(d). In other words, the $Z$ phase is adiabaticaly isomorphic to the collinear triple-$q$ state. Namely, the spin-longitudinal $S_z$ component exhibits the long-range order associated with the triangular-superlattice formation, while the spin-transverse ($S_x,S_y$) components are disordered on long length scale.
\begin{figure}[t]
	\centering
	\begin{tabular}{c}
		\begin{minipage}{\hsize}
			\includegraphics[width=\hsize]{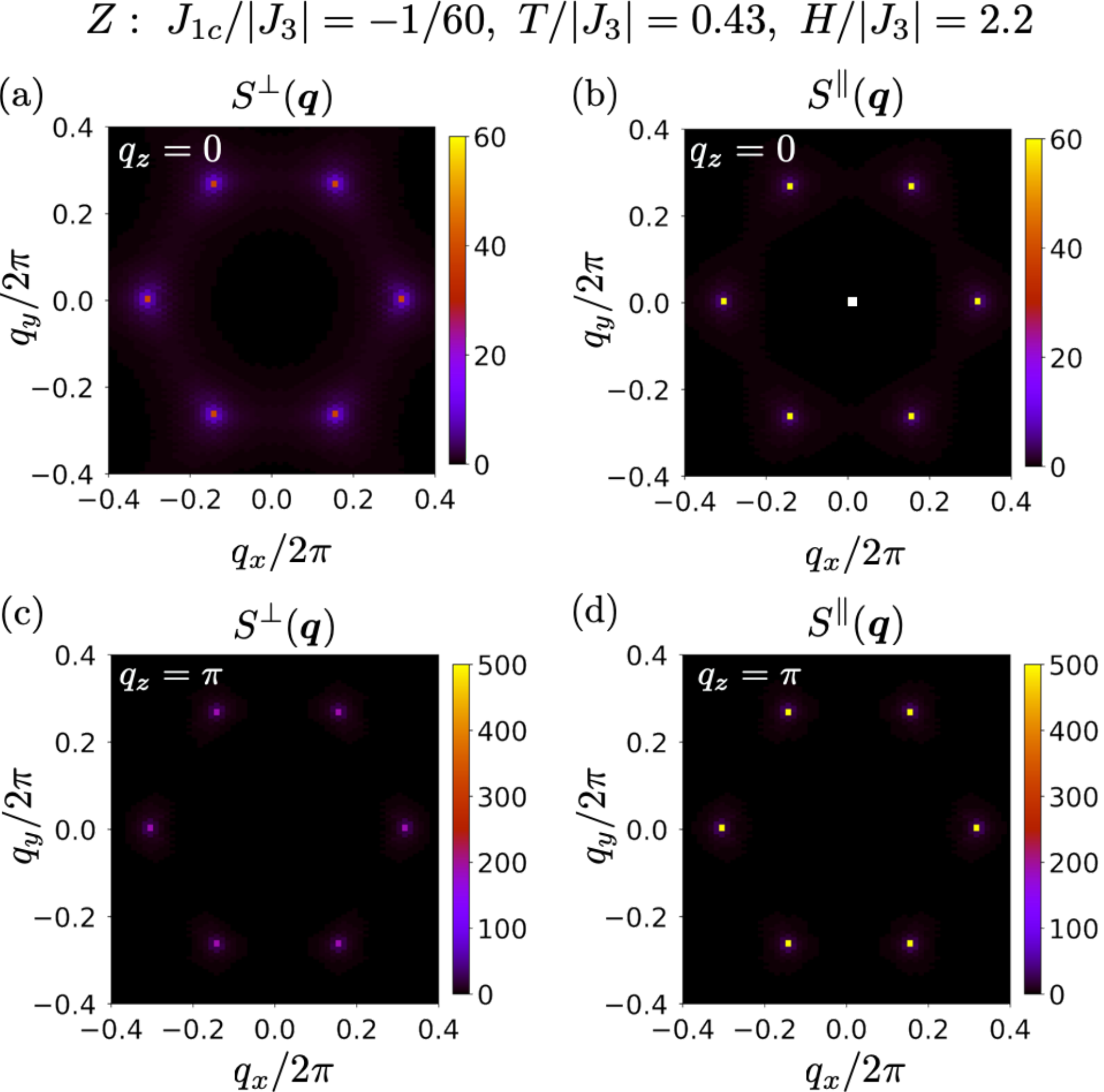}	
		\end{minipage}
	\end{tabular}
	\caption{
Spin structure factors in the $Z$ state at $T/|J_3|=0.43$ and $H/|J_3|=2.2$  for the weaker antiferromagnetic nearest-neighbor interplanar coupling of $J_{1c}/|J_3|=-1/60$ in the $(q_x,q_y)$ plane,  (a,b) at $q_z=0$, and (c,d) at $q_z=\pi$, representing (a,c) the perpendicular $S^\perp (\bm{q})$, and (b,d) the parallel $S^\parallel (\bm{q})$. The lattice size is $90\times 90\times 30$ ($L=90)$. The data are taken by the $T$-annealing run.
	}
	\label{3DFPD}
\end{figure}

 Typical real-space spin configurations of the $Z$ state are shown in Fig. 17 for three successive triangular layers. 
 One can see from the figure that, as observed in the SkX state, the skyrmion core keeps a tendency to form the $ABABAB\cdots $-type stacking pattern where the skyrmion core of the next layer is located at the center position of the triangle formed by those of the original layer, leading to the 3D magnetic long-range order in the spin-longitudinal component. 
\begin{figure}[t]
	\centering
	\begin{tabular}{c}
		\begin{minipage}{\hsize}
			\includegraphics[width=\hsize]{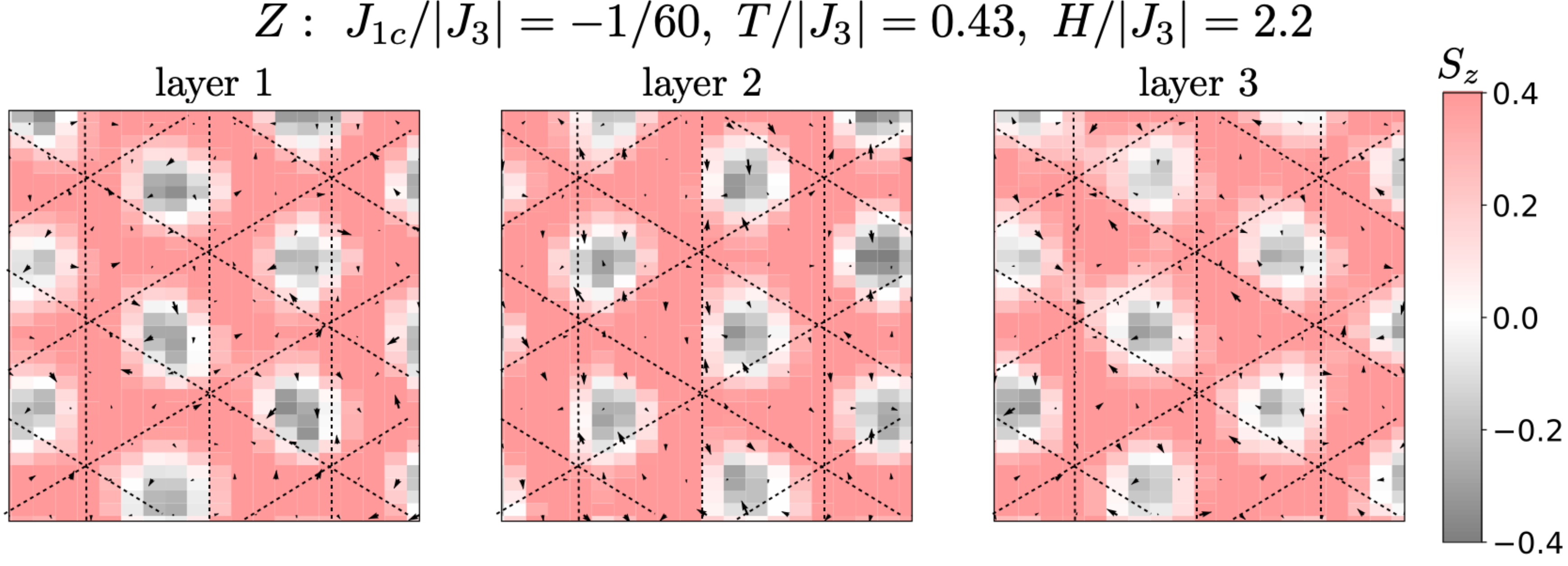}	
		\end{minipage}
	\end{tabular}
	\caption{
Typical real-space spin configurations in the $Z$ phase at  $T/|J_3|=0.43$ and $H/|J_3|=2.2$ for the weaker antiferromagnetic nearest-neighbor interplanar coupling of $J_{1c}/|J_3|=-1/60$ are shown for three successive triangular layers, from layer 1 to layer 3. The color represents the spin $S_z$ component, while the arrow represents the direction of the spin-transverse ($S_x,S_y$) components. To reduce the thermal noise, short-time averaging of 50 MCS is made, while the figure represents a part of the $90\times 90\times 30$ ($L=90)$ lattice with a common $xy$ section among layers 1-3. The data are taken by the $T$-annealing run.
	}
	\label{3DFPD}
\end{figure}
%
%

 As can be seen from Fig. 17, the spin-transverse ($S_x,S_y$) components are much reduced even after the short-time averaging of only 50 MCS, suggesting the transverse-spin disorder. Concerning the spatial distribution of the scalar chirality (not shown here), similarity to the case of the ferromagnetic $J_{1c}$, each triangular layer forms a random domain state consisting of finite-size SkX and anti-SkX domains within the layer, with a vanishing net scalar chirality even for each triangular layer. The mean size of these random SkX (anti-SkX) domains within the layer corresponds to the finite transverse spin correlation length in the triangular layer. The stacking pattern of such random-domain states in each layer looks random also along the stacking ($z$) direction without long-range correlations. 

 In Figs. 18, we show the $q_z$-dependence of (a) the perpendicular and (b) the parallel spin structure factors $S^\perp (q_x^*,q_y^*,q_z)$ and $S^\parallel (q_x^*,q_y^*,q_z)$ of the $Z$ phase computed by the $T$-exchange runs ($L\leq 72$) and by the $T$-annealing runs ($L\geq 78$). As can be seen from Fig. 18(a), $S^\perp (q_x^*,q_y^*,q_z)$ exhibits a peak at $q_z=\pi$. This $q_z=\pi$ peak exhibits a size crossover, i.e., its peak height tends to increase for smaller $L$, but decreases for the largest size $L=90$, suggesting that the $q_z=\pi$ peak of $S^\perp (q_x^*,q_y^*,q_z)$ might not be a truly divergent one.

 By contrast, as can be seen from Fig. 18(b), $S^\parallel (q_x^*,q_y^*,q_z)$ exhibits a $q_z=\pi$ peak with clear divergent $L_z$-dependence even including $L=90$, indicating the onset of the spin-longitudinal order with $q_z=\pi$. Meanwhile, a dull peak with its height systematically increasing with $L_z$ is observed also at $q_z=0$, though the observed $q_z=0$ peak is rather dull. Thus, it is not necessarily clear only from the present $S^\parallel (q_x^*,q_y^*,q_z)$ data whether the $q_z=0$ peak is divergent or not in the $L\rightarrow \infty$ limit.   

 Yet, Landau-type mean-field argument suggests that the $q_z=0$ peak of $S^\parallel (q_x^*,q_y^*,q_z)$ might also be a Bragg peak. The argument is as follows: Suppose that $\phi_z({\bm q})$ are the order-parameter fields of the $Z$ phase, where the subscript $z$ denotes the spin $S_z$ component. In magnetic fields which is essential for the stabilization of the $Z$ phase, a uniform-magnetization field $m_z=\phi_z({\bm 0})$ might also be important. In the Landau-type expansion given in terms of these order-parameter fields, the quartic term has a general form of $\phi_z({\bm q}_1)\phi_z({\bm q}_2)\phi_z({\bm q}_3)\phi_z({\bm q}_4)$ with a constraint ${\bm q}_1+{\bm q}_2+{\bm q}_3+{\bm q}_4=0$ (mod $2\pi$). Since the $Z$ state is the collinear triple-$q$ state, three out of four $\phi_z({\bm q}_j)$ should be ${\bm q}_1^*$, ${\bm q}_2^*$ and ${\bm q}_3^*$. Since ${\bm q}_{1,xy}^* + {\bm q}_{2,xy}^* + {\bm q}_{3,xy}^* = {\bm 0}$ by definition, the remaining fourth order-parameter field should have ${\bm q}_{4,xy}={\bm 0}$. The only possibility here is to use the uniform-magnetization field with ${\bm q}={\bm 0}$. This in turn entails the condition for the $q_z$-values of the other three order-parameter fields, $q_{1,z}+q_{2,z}+q_{3,z}=0$. Interestingly, the primary candidates of the order-parameter fields favored by the antiferromagnetic $J_{1c}$, $q_{j,z}=\pi$ for all $j=1,2,3$, cannot satisfy this constraint. By contrast, if one puts one out of three $q_{j,z}$ to be zero, this constraint can be satisfied. This argument certainly supports the appearance of the Bragg peak in $S_\parallel ({\bm q})$ not only at $q_z=\pi$ but also at $q_z=0$.  
\begin{figure}[t]
	\centering
	\begin{tabular}{c}
		\begin{minipage}{\hsize}
			\includegraphics[width=\hsize]{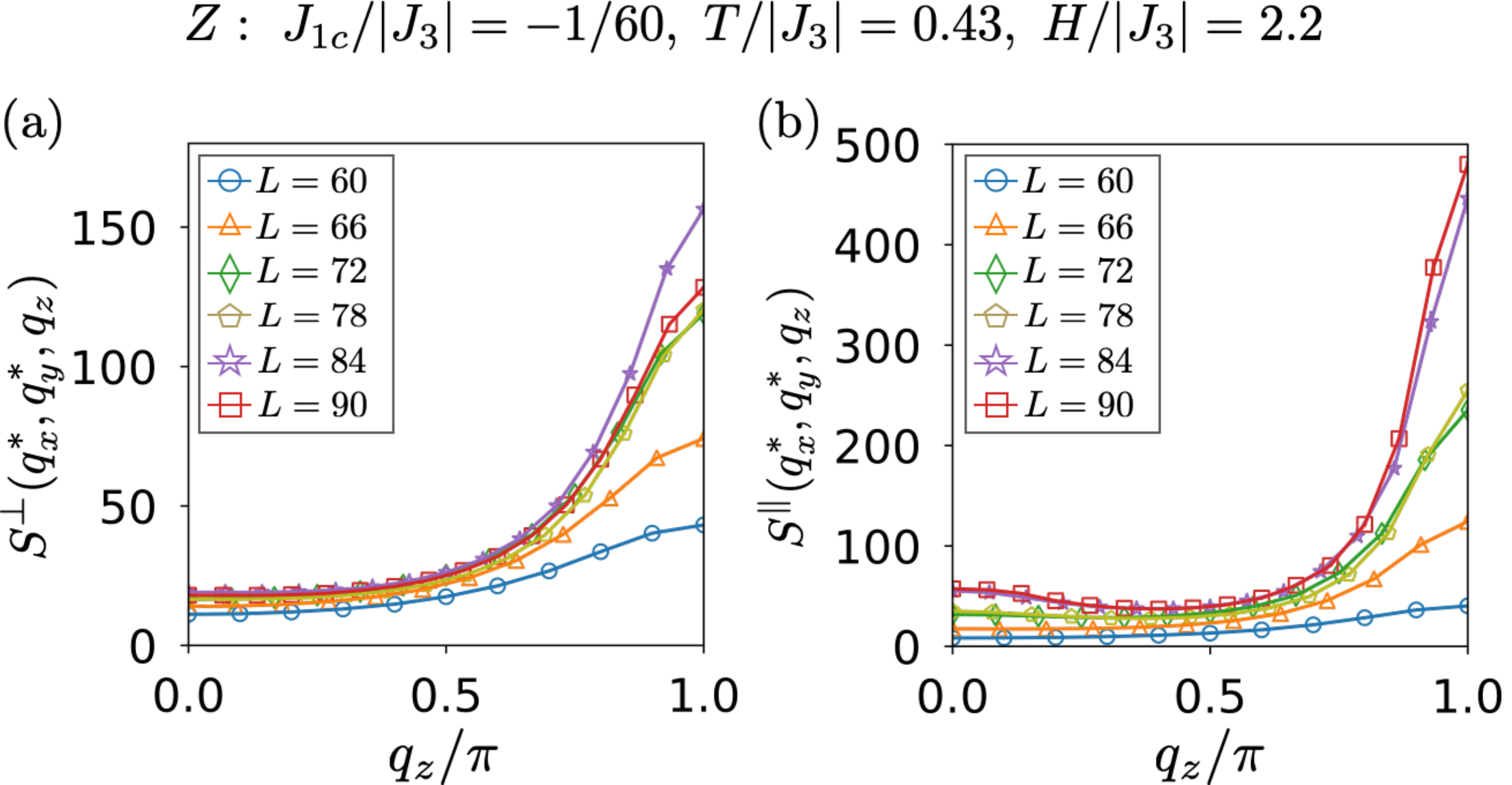}	
		\end{minipage}
	\end{tabular}
	\caption{
The $q_z$-dependence of the spin structure factors of (a) the spin-transverse components $S^\perp({\bm q})$, and of (b) the spin-longitudinal component $S^\parallel({\bm q})$ where $(q_x,q_y)$ is set to $(q_x^*, q_y^*)$, in the $Z$ phase at $T/|J_3|=0.43$ and $H/|J_3|=2.2$ for the weaker antiferromagnetic nearest-neighbor interplanar coupling of $J_{1c}/|J_3|=-1/60$.  The lattice sizes are  $L\times L\times \frac{1}{3}L$ with $L=60$, 66, 72, 78, 84 and 90.  The data for the sizes $L\leq 72$ are taken by the $T$-exchange runs, while the data for the sizes $L\geq 78$ are taken by the $T$-annealing runs.
	}
	\label{3DFPD}
\end{figure}
%

 Next, we move to the higher-$H$ region. In Fig. 19, we show the $T$-dependence of several physical quantities, including (a) the specific heat, (b) the total scalar chirality, (c) the transverse and (d) the longitudinal lattice $C_3$ symmetry-breaking parameters $m_3^\perp$ and $m_3^\parallel$ (the definition of $m_3^\perp$ and $m_3^\parallel$ have been given in Eqs. (4, 5) and (6, 7)), respectively, at a higher field of $H/|J_3|=4.0$ in the $T$ region involving the paramagnetic, single-$q$, double-$q$ and re-entrant single-$q$ phases. The behavior of these physical quantities are quite different between smaller sizes of $L\leq 66$ and larger sizes of $L\geq 72$. For smaller sizes of $L\leq 66$, the ordered state remains to be the single-$q$ conical-spiral state at any $T$, characterized by large $m_3^\perp$ and vanishing $m_3^\parallel$. In sharp contrast, for larger sizes of $L\geq 72$, a different state characterized by nonzero $m_3^\perp$ smaller than the value of the single-$q$ state, and by small but nonzero $m_3^\parallel$, sets in at $0.22 \lesssim T \lesssim 0.35$. As will be shown below, this state turns out to be the double-$q$ state. 

 These results indicate that the double-$q$ phase at higher fields is stabilized only for larger sizes of $L\gtrsim 72$, similarly to the case of the $Z$ phase which is also stabilized only for larger sizes of $L\gtrsim 72$. This probably reflects the fact that the double-$q$ phase is basically not favored by the antiferromagnetic $J_{1c}$, and even when the double-$q$ phase is eventually stabilized for the system with small enough $|J_{1c}|$ in the thermodynamic limit, relatively large sizes are required for such an asymptotic behavior to be visible.
\begin{figure}[t]
	\centering
	\begin{tabular}{c}
		\begin{minipage}{\hsize}
			\includegraphics[width=\hsize]{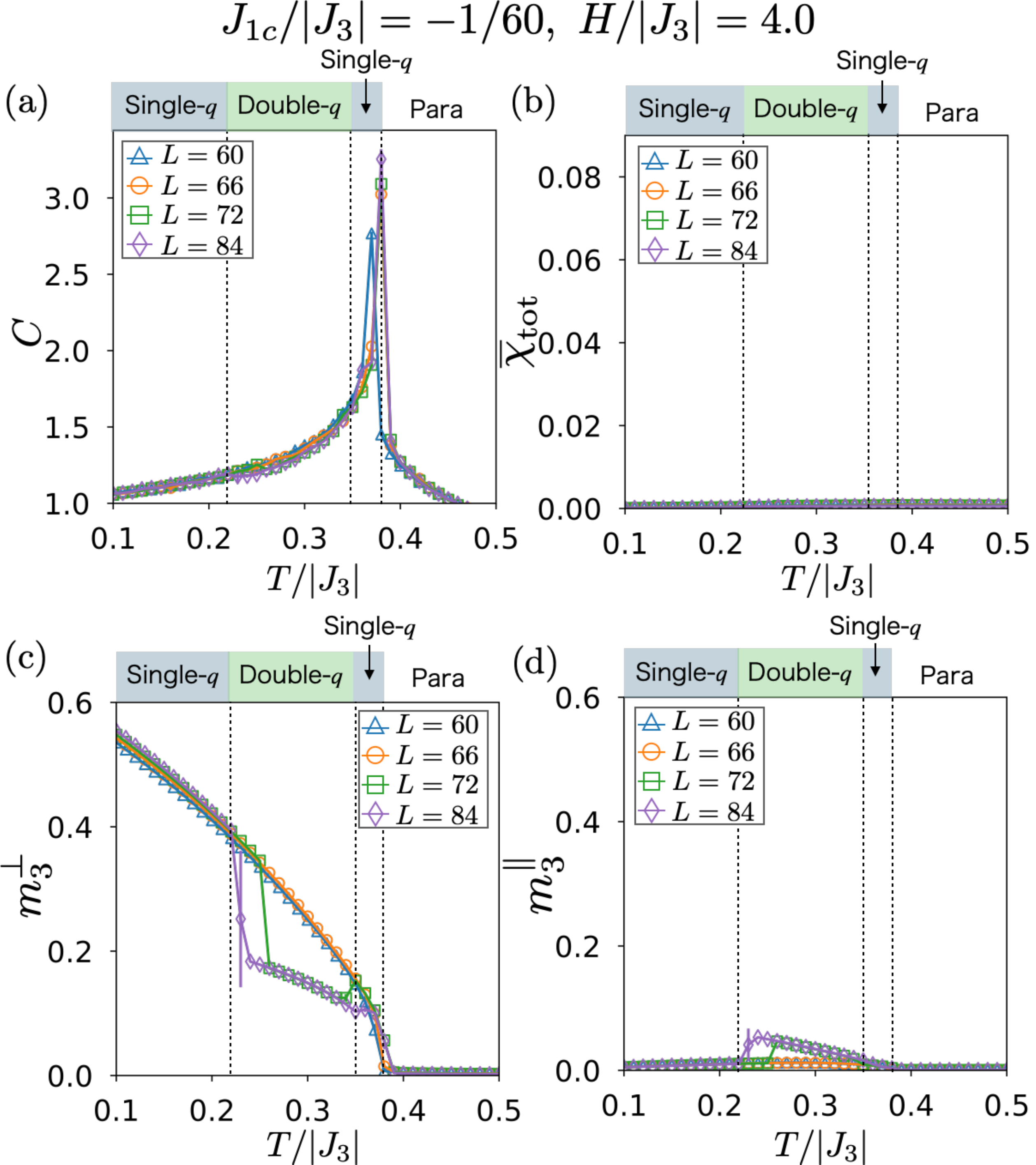}	
		\end{minipage}
	\end{tabular}
	\caption{
The temperature and size dependence of (a) the specific heat, (b) the total scalar chirality, (c) the transverse and (d) the longitudinal components of the lattice $C_3$-symmetry-breaking parameter, at a field $H/|J_3|=4.0$ for the weaker antiferromagnetic nearest-neighbor interplanar coupling of $J_{1c}/|J_3|=-1/60$. The lattice sizes are $L\times L\times \frac{1}{3}L$ with $L=60$, 66, 72 and 84. The data are taken by the $T$-annealing runs.
	}
	\label{3DFPD}
\end{figure}
%
%

 Typical spin structure factors $S(\bm{q})$ of the double-$q$ state are shown in the $(q_x,q_y)$ plane in Fig. 20, i.e., $S^\perp(\bm{q})$ and $S^\parallel (\bm{q})$ for $q_z=0$ in (a) and (b), and the ones for $q_z=\pi$ in (c) and (d), respectively. As can be seen from the figure, for both $q_z=0$ and $q_z=\pi$, $S^\perp(\bm{q})$ exhibits two pairs of peaks, while $S^\parallel (\bm{q})$ exhibits a single pair of peaks located at the complementary positions to those of $S^\perp(\bm{q})$. In the notation of Refs. \cite{Kawamura-review,Kawamura2024}, the state may be described as the ($2q,1q$) state, which is essentially of the same character as the one observed in the 2D model \cite{OkuboChungKawamura}.

 There is a big difference in the peak intensities of the transverse and the longitudinal components by about three orders of magnitudes. In $S^\perp(\bm{q})$, the $q_z=\pi$ component yields sharp Bragg peaks while the $q_z=0$ component yields very weak broader peaks corresponding to the short-range order. The situation is reversed in $S^\parallel(\bm{q})$, i.e., while the $q_z=0$ component yields sharp Bragg peaks, the $q_z=\pi$ component yields very weak broader peaks corresponding to the short-range order. 

 Concerning the reason why the double-$q$ Bragg peaks appear at $q_z=\pi$ whereas the single-$q$ Bragg peaks appear at $q_z=0$, one can employ Landau-type mean-field argument similar to the one employed above for the $S({\bm q})$ of the $Z$ phase. The order-parameter fields of the double-$q$ phase should include both the spin-transverse ($S_x,S_y$) components, ${\bm \phi}_{xy}({\bm q}_j^*)$ ($j=1,2,3$), and the spin-longitudinal $S_z$ components, $\phi_z({\bm q}_j^*)$ ($j=1,2,3$), where $q_{j,z}^*$ is either 0 or $\pi$. Since the double-$q$ phase is a high-field phase, magnetic field is expected to be important and a uniform-magnetization  field $m_z=\phi_z({\bm 0})$ would come into play. Remember that we are now interested in the interference of the spin-transverse and the spin-longitudinal components, i.e., the relation between  ${\bm \phi}_{xy}({\bm q}_j^*)$ and $\phi_z({\bm q}_j^*)$. The quartic term in the Landau-type expansion has such an interference term of the form ${\bm \phi}_{xy}({\bm q}_1)\cdot {\bm \phi}_{xy}({\bm q}_2)\phi_z({\bm q}_3)\phi_z({\bm q}_4)$, again with a constraint ${\bm q}_1+{\bm q}_2+{\bm q}_3+{\bm q}_4={\bm 0}$ (mod $2\pi$). By the definition of the double-$q$ state, transverse components should have ${\bm q}_1=({\bm q}_{1,xy}, \pi)$ and ${\bm q}_2=({\bm q}_{2,xy}, \pi)$ where we have assumed the primary contribution arising from $q_z=\pi$. For the two remaining longitudinal fields, one of them (the fourth one) might be a uniform field with ${\bm q}_4=({\bm 0}, 0)$. Then, the constraint ${\bm q}_1+{\bm q}_2+{\bm q}_3+{\bm q}_4={\bm 0}$ requires for the ${\bm q_3}$ of the remaining third field to be ${\bm q}_3=({\bm q}_{3,xy}^*,0)$. Hence, the Landau-type argument gives the reason why the longitudinal spin order yields a single-$q$ pair of Bragg peaks at the complementary position to the double-$q$ Bragg-peak positions of the spin-transverse components at $q_z=0$, rather than at $q_z=\pi$.

%
%
\begin{figure}[t]
	\centering
	\begin{tabular}{c}
		\begin{minipage}{\hsize}
			\includegraphics[width=\hsize]{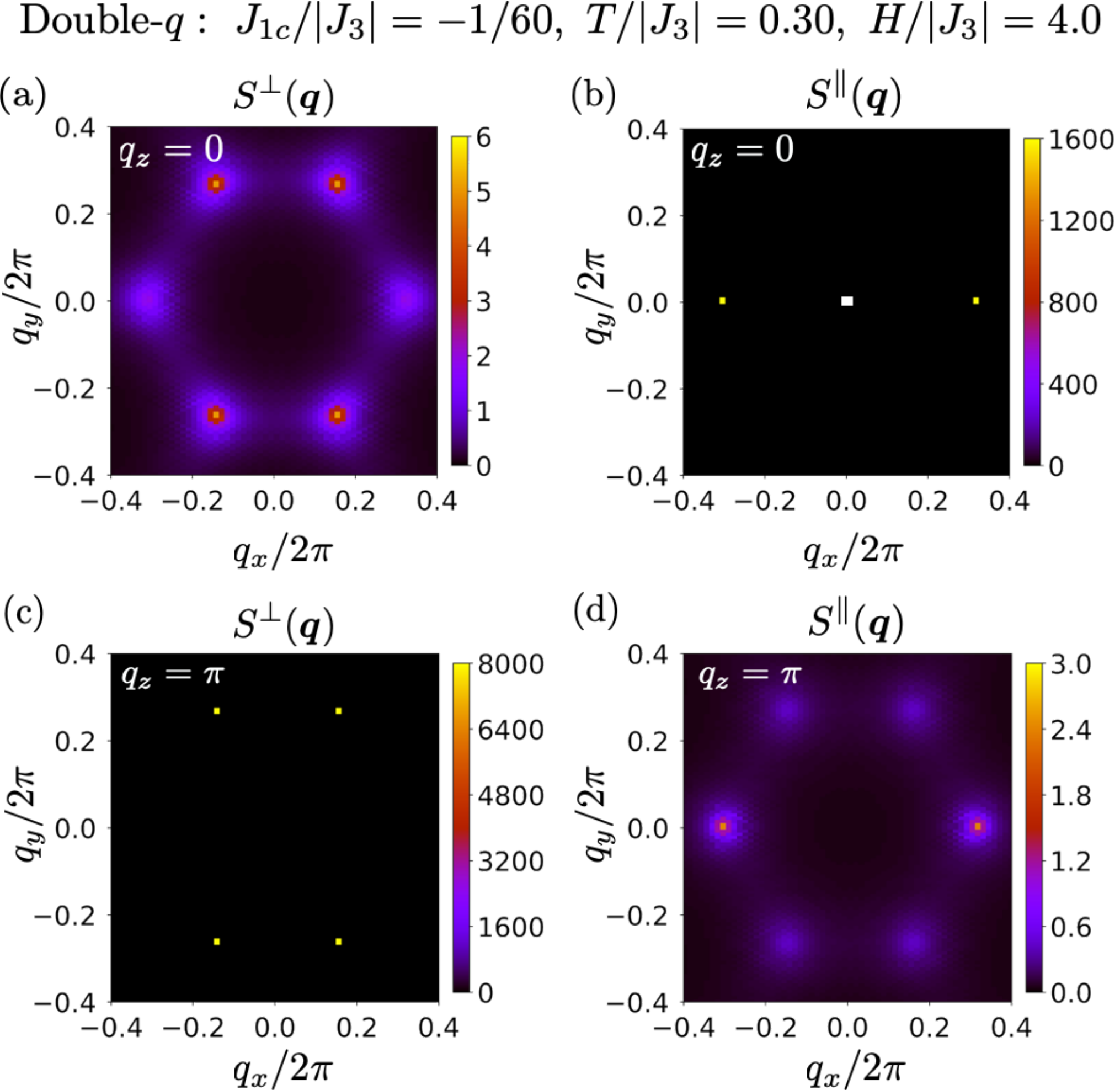}	
		\end{minipage}
	\end{tabular}
	\caption{
Spin structure factors in the double-$q$ phase at $T/|J_3|=0.30$ and $H/|J_3|=4.0$  for the weaker antiferromagnetic nearest-neighbor interplanar coupling of $J_{1c}/|J_3|=-1/60$ in the $(q_x,q_y)$ plane,  (a,b) at $q_z=0$ and (c,d) at $q_z=\pi$, representing (a,c) the perpendicular $S^\perp (\bm{q})$ and (b,d) the parallel $S^\parallel (\bm{q})$. The lattice size is $72\times 72\times 24$ ($L=72$). The data are taken by the $T$-annealing run.
	}
	\label{3DFPD}
\end{figure}

\section{Summary and discussion}

 The nature of the magnetic ordering and the $T$-$H$ phase diagram of the frustrated isotropic Heisenberg model on a stacked-triangular lattice is investigated by extensive MC simulations for both cases of the ferromagnetic  and the antiferromagnetic nearest-neighbor interplanar couplings $J_{1c}$, in order to clarify the effects of the three-dimensionality (interplanar coupling) on centrosymmetric SkX formation.

 The SkX phase turns out to be stabilized at finite fields and at finite temperature for both ferromagnetic and antiferromagnetic $J_{1c}$, together with the $Z$ phase which is a random domain state consisting of both SkX and anti-SkX domains. While the SkX state is robust against the change of the strength of the interplanar coupling $J_{1c}$ for ferromagnetic $J_{1c}$, it is easily destabilized by modestly weak antiferromagnetic $J_{1c}$. Indeed, the magnetic phase diagram of the 3D short-range model with moderate or strong antiferromagnetic $J_{1c}$  consists of only the single-$q$ spiral phase. Meanwhile, the magnetic phase diagrams of the 3D short-range model with the ferromagnetic or sufficiently weak antiferromagnetic $J_{1c}$ turn out to be similar to those of the 2D short-range \cite{OkuboChungKawamura} or the 2D long-range RKKY \cite{MitsumotoKawamura2022} models, involving the field-induced SkX phase. It also turns out that the RSB phenomenon observed in the 3D long-range RKKY model \cite{MitsumotoKawamura2021} is not realized in the 3D short-range model, suggesting that the long-range nature and/or the stronger frustration along the interplanar direction inherent to the RKKY interaction might be important for the occurrence of the RSB.

 The stacking pattern of SkX layers along the direction perpendicular to the triangular layers is also studied. In the case of ferromagnetic $J_{1c}$, the stacking pattern of SkX layers is a direct on-top stack along the stacking direction where the spin configuration is uniform along the stacking direction. The resulting 3D SkX is simply a triangular superlattice of skyrmion tubes, the associated spin structure factors $S({\bm q})$ exhibiting triple-$q$ Bragg peaks at (${\bm q}_{j,xy}^*,0$) ($j=1,2,3$). Such a 3D SkX state possesses the net total scalar chirality, resulting in the eminent topological Hall effect in bulk 3D systems. Reflecting the $Z_2$ chiral degeneracy of the present model without the dipolar or the spin-orbit couplings, the total scalar chirality could equally be negative or positive, leading to either the SkX state or the anti-SkX state as a consequence of spontaneous $Z_2$-symmetry breaking. In real magnets, the residual dipolar interaction or the spin-orbit interaction might weakly break the $Z_2$ symmetry, and the SkX state (or the anti-SkX state depending on the situation) might be chosen as an ordered state.

  In the case of antiferromagnetic $J_{1c}$, the stacking pattern of SkX layers is no longer a direct on-top stack, but rather a slided stack along the stacking direction. In fact, many stacking configurations of SkX layers are  energetically degenerate. In thermal equilibrium, the stable stacking pattern turns out to be such that the SkX core of a given SkX layer is located at the center position of the triangle formed by the skyrmion cores in the adjacent SkX layer, with the resulting 3D SkX forming a $ABABAB\cdots $-type hcp-like stacking pattern along the stacking direction. Such a stacking pattern of SkX layers is likely to be stabilized not just by energetical reason but also by entropical reason via the order-from-disorder mechanism \cite{Villain,KawamuraSW,Henley}. The associated spin structure factors $S({\bm q})$ exhibit Bragg peaks both at (${\bm q}_{j,xy}^*, \pi$) and at (${\bm q}_{j,xy}^*, 0$) ($j=1,2,3$).

 We note that, reflecting the energetical degeneracy among many stacking patterns of SkX (anti-SkX) layers, when the full equilibration is not achieved in MC simulations, various other stacking patterns of SkX (anti-SkX) layers also appear. In fact, in the layer stacking consisting solely of SkX-layers (of anti-SkX-layers), arbitrary combinations of $(A,B,C)$ SkX (anti-SkX) positioning in each layer are all energetically degenerate so long as the same SkX (anti-SkX) positioning, e.g., $AA$, is inhibited in the adjacent layers. Note that these nontrivial slided stacking and the associated heavy degeneracy are peculiar to the antiferromagnetic $J_{1c}$, and do not arise for the ferromagnetic $J_{1c}$. In spite of such difference in the stacking patterns and the associated spin configurations of the antiferromagnetic $J_{1c}$ from those of the ferromagnetic $J_{1c}$, the resulting 3D SkX state still keeps the net total scalar chirality irrespective of the signs of $J_{1c}$, resulting in the eminent topological Hall effect in bulk.

 Furthermore, in some cases, quenching the system from high $T$ often leads to the metastable SkX state consisting of random stacking of both SkX layers and anti-SkX layers. Although such states have energies slightly higher than the energy of the state consisting of all SkX layers (or of all anti-SkX layers), they remain metastable once generated, and might exhibit the suppressed or vanishing topological Hall effect because of the cancellation of the Hall signal between the SkX-layers part and the anti-SkX-layers part. In real magnets, however, weak perturbative interactions breaking the $Z_2$ chiral degeneracy, i.e., the dipolar or the spin-orbit couplings, would energetically bias the one from the other, leading to the net topological Hall effect. 

 One may call these randomly-stacked metastable states peculiar to tne antiferromagnetic $J_{1c}$ ``skyrmion-layer glass''.  
Such skyrmion-layer glass states, though they are metastable states, accompany slow dynamics, leading to interesting non-equilibrium glassy behaviors. 

 For the case of the antiferromagnetic $J_{1c}$,  we have confined ourselves in the present paper to the two specific values of $J_{1c}/J_3$ corresponding to two distinct typical ordering behaviors, i.e., $J_{1c}/J_3=-\frac{1}{60}$ and $J_{1c}/J_3=-\frac{1}{15}$. Meanwhile, how the ordering behavior and the magnetic phase diagram change in the $J_{1c}$-region between these two $J_{1c}$-values is not quite clear and remains as an open problem. In particular, on increasing the $|J_{1c}|/J_3$ value from $\frac{1}{60}$ toward $\frac{1}{15}$, how the nontrivial multiple-$q$ phases, i.e., the triple-$q$ SkX phase, the $Z$ phase and the double-$q$ phase, go away from the phase diagram is an interesting open question. Numerically, however, clarifying this might require considerable amount of computational efforts, not only because the ordering behavior and the phase diagram could be rather complex in this crossover region, but also because, the information of larger lattice sizes might become necessary in order to reveal the asymptotic bulk ordering behavior in this region, as we have already seen for the case of $J_{1c}/J_3=-\frac{1}{60}$.

 Not just limited to the interplanar coupling $J_{1c}$, possible dependence of the phase diagram on the intraplanar couplings $J_1,J_2,J_3, \cdots$ could also be the issue. Ref. \cite{OkuboChungKawamura} indicated that in 2D the qualitative features of the phase diagram were basically common between the $J_1-J_3$ and the $J_1-J_2$ models. Natural expectation then would be that the qualitative features of the phase diagram are rather insensitive to the details of the interplanar couplings so long as the interplanar coupling $J_{1c}$ is limited to nearest neighbors (the introduction of the frustrated further-neighbor interplanar couplings sometimes modifies the ordering behavior considerably \cite{LinBatista3D}).

 Finally, we wish to discuss the possible relevance of the present results to experiments on centrosymmetric SkX-hosting magnets including, e.g., triangular Gd$_2$PdSi$_3$ \cite{Kurumaji,Hirschberger2020PRL}, breathing-kagome Gd$_3$Ru$_4$Al$_{12}$ \cite{Hirschberger2019} and tetragonal GdRu$_2$Si$_2$ \cite{KhanhSeki2020} and EuAl$_4$ \cite{Takagi2022}. In all of them, the SkX state is stabilized at finite fields, while the observed SkX structures are triangular in the former two, and square in the latter two.  All of them are metallic magnets consisting of Gd$^{3+}$ (Eu$^{2+}$) localized Heisenberg spins interacting via the long-range RKKY interaction which oscillates in sign leading to the magnetic frustration, where the eminent topological Hall effect has been observed in common. The spin configurations in the SkX state are all uniform along the $z$-direction, i.e., $q_z^*=0$.

 To the authors' knowledge, there was no report of the $q_z^*=\pi$ Bragg peak in the SkX state. Since the SkX-hosting metallic compounds quoted above possess the long-range RKKY interaction, and the numerical calculation on the 3D RKKY model often yields the $q_z^*=0$ SkX order \cite{MitsumotoKawamura2021}, the experimental result seems natural and quite likely.

 Yet, it might be interesting to seek for the SkX-hosting magnet with predominantly antiferromagnetic interplanar coupling. Such SkX would also exhibit the topological Hall effect as observed in other SkX-hosting compounds, at least in thermal equilibrium. The eminent characteristic of the SkX state with the antiferromagnetic $J_{1c}$ would be that, in addition to the Bragg peaks located at $q_z^*=0$, the dominant Bragg peaks also appear at $q_z^*=\pi$.

 In the SkX state with predominantly antiferromagnetic interplanar coupling, the way of stacking of SkX layers is heavily degenerate unlike the case of the predominantly ferromagnetic interplanar coupling. Such degeneracy might lead to interesting glassy behaviors.

 We also find that the RSB, which arises in the 3D long-range RKKY model, does not arise in the 3D short-range $J_1-J_3-J_{1c}$ model. The result suggests that both the three-dimensionality and the long-range nature of interaction is necessary to realize the RSB in the centrosymmetric SkX state. Yet, since most of the experimental centrosymmetric SkX states are identified so far in metallic magnets interacting via the long-range RKKY interaction with non-negligible interplanar couplings \cite{Kawamura-review}, there seems to be a good chance of experimentally observing intriguing RSB phenomena realized in the 3D centrosymmetric SkX states.     

 This study was supported by JSPS KAKENHI Grants No.17H06137 and No.24K00572. We are thankful to ISSP, the University of Tokyo, for providing us with CPU time.

\end{document}